\documentclass[aps,pra,superscriptaddress,nofootinbib,floatfix,preprint]{revtex4}

\usepackage{graphicx}	

\newcommand{\beq}{\begin{equation}}
\newcommand{\eeq}{\end{equation}}
\newcommand{\beqa}{\begin{eqnarray}}
\newcommand{\eeqa}{\end{eqnarray}}
\newcommand{\bpr}{\begin{problem}}
\newcommand{\epr}{\end{problem}}
\newcommand{\bcent}{\begin{center}}
\newcommand{\ecent}{\end{center}}
\newcommand{\bfig}{\begin{figure}}
\newcommand{\efig}{\end{figure}}
\newcommand{\bpc}{\begin{picture}}
\newcommand{\epc}{\end{picture}}
\newcommand{\barr}{\begin{array}}
\newcommand{\earr}{\end{array}}
\newcommand{\bitm}{\begin{itemize}}
\newcommand{\eitm}{\end{itemize}}
\newcommand{\bright}{\begin{flushright}}
\newcommand{\eright}{\end{flushright}}
\newcommand{\bminip}{\begin{minipage}}
\newcommand{\eminip}{\end{minipage}}
\newcommand{\btab}{\begin{tabular}}
\newcommand{\etab}{\end{tabular}}

\newcommand{\nnb}{\nonumber}

\newcommand{\MP}{M_{\rm P}}

\newcommand{\hiroshima}{Graduate School of Science, Hiroshima University, Kagamiyama, Higashi-Hiroshima 739-8526, Japan}
\newcommand{\lmu}{Ludwig-Maximilians-Universit$\ddot{a}$t M$\ddot{u}$nchen, Fakult$\ddot{a}$t f. Physik, Am Coulombwall 1, D-85748 Garching, Germany}

\newcommand{\IZEST}{IZEST}
\newcommand{\bra}{<\hspace{-.2em}}
\newcommand{\ket}{\hspace{-.2em}>}
\newcommand{\dbra}{<\hspace{-.45em}<}
\newcommand{\dket}{>\hspace{-.42em}>}

\begin{document}
\title{
Fundamental Physics Explored with High Intensity Laser
}
\author{T. Tajima} \affiliation{\IZEST}\affiliation{\lmu}
\author{K. Homma} \affiliation{\IZEST}\affiliation{\hiroshima}
\date{September 11, 2012}
\begin{abstract}
Over the last Century the method of particle acceleration to high energies has become the prime approach to explore the fundamental nature of matter in laboratory. It appears that the latest search of the contemporary accelerator based on the colliders shows a sign of saturation ( or at least a slow-down ) in increasing its energy and other necessary parameters to extend this frontier. We suggest two pronged approach enabled by the recent progress in high intensity lasers. First we envision the laser-driven plasma accelerator may be able to extend the reach of the collider. For this approach to bear fruit, we need to develop the technology of high averaged power laser in addition to the high intensity.  For this we mention that the latest research effort of ICAN is an encouraging sign. In addition to this, we now introduce the concept of the non-collider paradigm in exploring fundamental physics with high intensity (and large energy) lasers. One of the examples we mention is the laser wakefield acceleration (LWFA) far beyond TeV without large luminosity. If we relax or do not require the large luminosity necessary for colliders, but solely in ultrahigh energy frontier, we are still capable of exploring such a fundamental issue. Given such a high energetic particle source and high-intensity laser fields simultaneously, we expect to be able to access new aspects on the matter and the vacuum structure from fundamental physical point of views. LWFA naturally exploits the nonlinear optical effects in the plasma when it becomes of relativistic intensity. Normally nonlinear optical effects are discussed based upon polarization susceptibility of matter to external fields. We suggest application of this concept even to the vacuum structure as a new kind of order parameter to discuss vacuum-originating phenomena at semi-macroscopic scales. This view point unifies the following observables with the unprecedented experimental environment we envision; the dispersion relation of photons at extremely short wavelengths in vacuum (a test of the Lorentz invariance), the dispersion relation of the vacuum under high-intensity laser fields (non-perturbative QED and possibly QCD effects), and wave-mixing processes possibly caused by exchanges of low-mass and weakly coupling fields relevant to cosmology with the coherent nature of high-flux photons (search for light dark matter and dark energy). These observables based on polarization susceptibility of vacuum would add novel insights to phenomena discovered in cosmology and particle physics where order parameters such as curvature and particle masses are conventionally discussed. In other words the introduction of high intensity laser and its methodology enriches the approach of fundamental and particle physics in entirely new dimensions.
\end{abstract}

\maketitle
\section{Introduction}
  The method of exploration of the subatomic structure initiated by 
Rutherford has been so successful because it fits beautifully with 
the fundamentals of quantum mechanics. According to the Heisenberg 
uncertainty principle, the more minute the size of 
resolution in unraveling the structure of matter is demanded, the 
higher the momentum of the penetrating test particle
momentum should be. Thus the last Century saw the birth and spectacular
flourish of high energy (and high momentum) physics. With ever 
higher energies we have been able to explore ever finer details 
of matter's inner structure, discovering hierarchical layers of 
new particles with higher energies.

  In order to continue this quest, the currently preferred approach 
is that of the collider in which two opposing beams of charged 
particles are collided (e.g. \cite{MTigner}). 
The collider has the advantage in extending the momentum of the 
probing beam linearly proportional to the energies of each of 
the two beams, as opposed to the previous generation of fixed target 
accelerator, in which the probing momentum increases only as a 
function of square-root of energy. The linear accelerator size is 
determined by the accelerating gradient (among other factor). 
On the other hand, the accelerating gradient of the contemporary 
accelerator is limited by the metallic surface breakdown of the 
accelerating structure, i.e. the waveguide surface or more precisely 
at the tip of the slow-wave structure (say $\sim 100$MeV/cm = eV/Angst 
typical of the Keldysh breakdown field\cite{Keldysh}, 
but in reality, it is much 
less than this such as O$(10^2)$ MeV/m due to the material imperfection).
This is the first and main reason why a linear collider at a TeV frontier
extends over multi 10's km. The laser wakefield accelerator 
(LWFA, \cite{TT-Dawson}) is based on this recognition that in order to 
gain orders of magnitudes leap in the accelerating gradient, 
one has to employ a material that is already broken down and 
cannot break down further, i.e. that of plasma. Subsequent
derivatives such as wakefield acceleration driven by electron 
beams\cite{Chen} and by ion beams \cite{Caldwell} are based 
on the same idea. The plasma allows essentially arbitrary large 
accelerating gradient, primarily limited by the driver 
\cite{Zheng2012}, such as TeV/cm. However, the collider requires 
high luminosity
\beq\label{eq1}
{\cal L} = N^F_1 N^F_2 f / A,
\eeq
where $N^F_i$ is the number of Fermionic particles in the $i$-th beam and 
$f$ and $A$ are the repetition rate of the beam and 
the area of the focused beam. 
Unlike the fixed target accelerator where the fixed target may have the 
density of solid, the collider has to explore the opposing beam as 
a target whose density (and thus the total number of particles) is 
much more modest. As a general rule of thumb, the cross section of 
a particle to be newly discovered is inversely proportional to the 
square of the energy, as explained earlier through the Heisenberg 
principle. Thus we need to increase the luminosity of a collider much 
faster than the energy, as the collider is the main avenue of
high-energy accelerator physic\cite{Chao-Tigner}. 
This means that the figure of merit, the product of energy and luminosity,
is proportional to the cube of energy. Since very crudely the cost 
of an accelerator may be proportional to this product, we face an even 
taxing cost, as well as an ever escalating challenge to increase 
the load of more particles in the accelerator units.

So, the challenge laid down in front of us is how to meet these 
difficulties. We first respond to this challenge in reducing 
the size of accelerators by increasing the accelerating gradient 
by orders of magnitude by the plasma acceleration mentioned already. 
LWFA is capable of reducing the size of accelerators, in principle, 
by orders of magnitude. Beam-driven wakefields (as long as beams are 
generated by the conventional accelerators) are capable of reducing 
the size of them by some modest amount and / or converting proton 
energies into electron ones (in the case of proton driven wakefield).
In all cases here a serious consideration of luminosity awaits us. 
We will mention one such an effort toward addressing this question.

We then present our ideas on the non-collider paradigm. 
Once we remove the collider luminosity requirement, 
we no longer address many of particle physics issues that 
demand the discovery of (very mini-scale) high energy particles. 
On the other hand, by removing the demanding luminosity 
condition, perhaps we can concentrate on other parameter(s) to leap 
to an unprecedented frontier level. For example, Suzuki challenged 
us \cite{SLAC-lec.-2008} to look for a possibility to reach for PeV. 
Although we regard a PeV collider not feasible with the accompanying 
luminosity requirements, the LWFA might be capable of attaining its 
energy itself. If such high energies are valuable in 
exploring fundamental physics questions that do not require luminosity,
can we march toward such a goal?  

The current two approaches in particle collider physics and cosmology
have a huge gap in the spacetime scales of exploration. 
Particle physics is based on massive particle states
produced from the excited vacuum at the microscopic scale.
Meanwhile, cosmological observations are based on the curvature of
the vacuum at the macroscopic scale.
How can we introduce a unified view point to these different observables 
at extremely different scales? Here the history of laser may be 
instructive to learn how laser unveiled the matter structure at 
different scales. The structure of atom, the basic element of 
matter, has been probed by the particle beam probes with shorter wavelengths
than the size of atoms. After the invention of laser
we found semi-macroscopic collective phenomena by stimulating 
the collection of atoms by coherent laser photons. Note here that 
the individual wavelength of laser photons is much longer than that of atoms. 
We may see an analogy that 
as if particle physics is investigating local responses of the vacuum
corresponding to the ionization process of atoms, 
{\it i.e.}, microscopic polarization processes by producing real fermion pairs,
as we see in the production of high-momentum di-quark jets at 
electron-positron collision points. 
Meanwhile in cosmology we discuss only a bulk shape of the vacuum. 
These observations in two extreme scales
do not directly unveil phenomena which may appear only at semi-macroscopic 
scales of the spacetime. What happens if we stimulate the vacuum by 
high-intensity coherent photons in semi-macro scales between the two extreme
scales ? If unknown spacetime elements have couplings 
to photons, we might be able to track down 
unique semi-macroscopic phenomena in the
vacuum\cite{DEapb}. Based on this view point as the analogy to atomic structure
of matter, it is natural to try to apply the polarization susceptibility 
even to define unique observables for 
the semi-macroscopic structure of the vacuum. 
The polarization ${\cal P}$ in matter under laser irradiation is expressed as 
\beq\label{eq0}
{\cal P}(\omega) = \chi_1(\omega) E(\omega) + \chi_2(\omega)E^2(\omega) + \chi_3(\omega)E^3(\omega) + \cdot\cdot\cdot,
\eeq
where $\omega$ is frequency of laser photons, $E(\omega)$ is the laser 
electric field, and $\chi_i(\omega)$ are susceptibilities for respective 
order $i$. 

If we apply this treatment even to the vacuum, the constancy of the
velocity of light in vacuum, Lorentz invariance, means
that $\chi_1$ does not depend on $\omega$. 
If we envision the generation of very high energy photons 
via radiations from electrons by LWFA , we can probe whether 
this susceptibility is truly constant or not over a wide range 
of photon energy even reaching PeV as in section III.

We now focus on the second order term. The $\chi_2$ term corresponds to 
the photon-photon interaction in the vacuum. 
We can explore the $\gamma$photon-laser interaction in 
addition to the laser-laser interaction. We discuss the refractive 
index of photon with different $\omega$ under high-intensity laser fields 
by measuring both the real and imaginary parts of the refractive index 
of laser induced vacuum and these dispersion relations\cite{Shore}. 
We may also try to measure second harmonic
generation via photon-laser interactions to probe $\chi_2$. The cause of these 
phenomena are supposed to be due to the nonlinear QED effect and possibly the 
nonlinear QCD effect depending on $\omega$.
We may rather simply treat these phenomena as a part of quantum features of 
photon itself. However, once photons form a collective state such as a coherent
state in a laser field, especially the imaginary part may be treated 
as if the tunneling process occurs in the Dirac sea, as we will discuss in 
section IV. In this view point, the vacuum state is already treated as a part 
of matter in a sense. On the other hand, if there are additional effects 
other than these known non-perturbative processes from QED and possibly QCD, 
we are already investigating $\chi_2$ of more general fields\cite{APB-QED}. 

In principle we expect that even the third order effect could happen. 
If we mix more than two photon beams at a time, for example, we can try to 
measure optical parametric effects, yielding frequency sum or difference 
processes, if the vacuum contains nonlinearity whatever the origin
of such an interaction may be. 
It is important to note that the interaction rate should be cubic 
of the number of photons contained in the beams. Therefore, the sensitivity to 
probe $\chi_3$ should leap, if laser beams with a large number of photons
are available. This implies that extremely feeble $\chi_3$
may be measurable, that is, the weakly coupling of 
the new fields to the collective photons may be probed. 
Such new fields may 
correspond to Dark Matter and Dark Energy in the expression of the contemporary
cosmology. As long as we measure $\chi_3$ at the optical frequency, 
the explorable mass range of the Dark Matter and Dark Energy candidates 
naturally reaches sub-eV mass range. The particle collider 
paradigm yields interaction rates quadratic of the number of charged 
particles, typically $10^{11}$ particles per bunch. Compared to this,
the cubic nature of $\chi_3$ processes leads to a huge advantage of this 
proposed methodology, as we foresee more than Avogadro's number per 
laser pulse for the near future experiments\cite{HommaISMD}.  
We discuss this example in section V.

\section{Laser Acceleration for a Compact Collider}
\subsection{Ultrafast Intense Laser for High-Energy Acceleration}
\begin{figure}
\begin{center}
\includegraphics[width=0.8\linewidth]{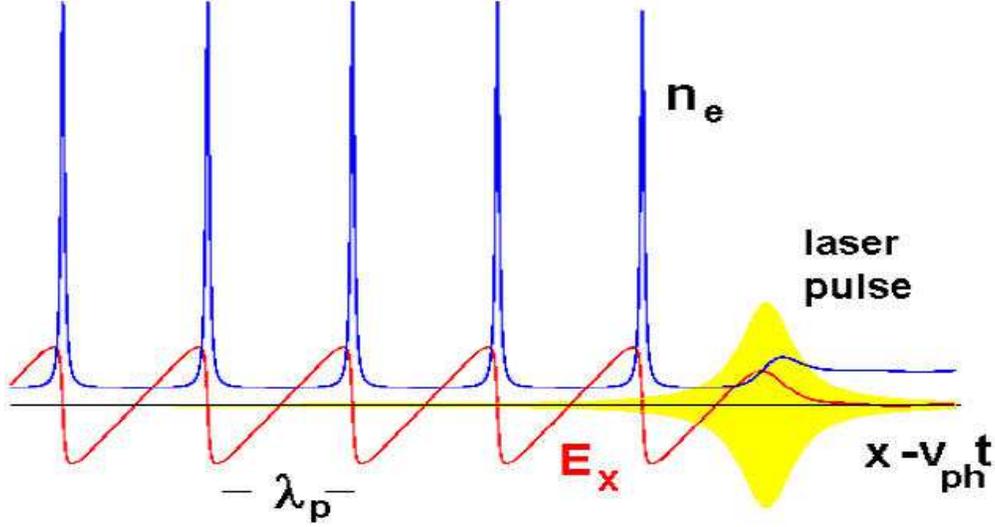}
\caption{
In laser wakefield generation, the property of the laser inducement 
of electronic plasma wave is such that the electron response to catch up with 
the laser's relativistic phase make the electron wave coherent and steepens, 
as seen in this 1D particle simulation picture. Since all electrons can do is 
speed up to $c$ so that the electron density piles up to from a cusp 
singularity.  As a  result, the large accelerating gradient on the order of 
Tajima-Dawson field (in fact in this example, exceeding that value) is 
generated that speeds with the laser pulse. Because of this relativistic 
coherence, the wakefield can be robust, unlike in the nonrelativistic 
large amplitude waves that tend easily wave-break\cite{Tajima2011}.
}
\label{Fig1}
\end{center}
\end{figure}

  The leaser wakefield acceleration (LWFA) is capable of creating
an accelerating gradient a few orders of magnitude beyond the 
conventional technology. Its accelerating gradient is on 
the order of magnitude $E_0 = m_e\omega_{pe} c / e$, the Tajima-Dawson 
field\cite{TT-Dawson}, typically on the order of GeV/cm,
where $\omega_{pe}$ is the plasma frequency.
In order to harness this large accelerating gradient in plasma 
that is a broken down matter and thus does not seem to hold any 
organized structure within itself, we introduce two principles, 
the collective acceleration and what we call the relativistic coherence.
The former was first considered by Veksler\cite{Veksler} in which plasma 
is excited by an incoming beam of electrons. Some attempts have been 
carried out to harness collective fields of plasma for acceleration 
\cite{Rostoker-Reiser}. However, stable phase ({\it i.e.} trapping) 
relationship in the acceleration of particles with the excited 
plasma waves has never been established
\footnote{ 
The first laser acceleration of protons from the sheath of electrons 
that was driven by laser from the back of the target had brought 
in the revisit of this phase stability issue in a new reincarnation
\cite{Snavely2000}; see review\cite{Tajima2010}.
}.
The advent of laser wakefield acceleration started by Tajima and 
Dawson in 1979\cite{TT-Dawson} made the exploitation of the 
collective fields of plasma for intense acceleration a subject 
of contemporary interest. In LWFA an prescribed laser pulse in 
Tajima and Dawson is capable of exciting a robust and large 
accelerating structure by the plasmas collective oscillating 
resonance, i.e. the plasma wave that is characterized by the plasma 
oscillation frequency $\omega_{pe}$ and its associated wavenumber 
$k_p = \omega_{pe}/ c$.
Further, Tajima and Dawson recognized the latter, the relativistic 
coherence, in exciting plasma oscillations in the relativistic regime 
in which the wakefield driving entity, the tailored laser pulse 
which moves at a relativistic speed $v_g$ 
(the group velocity of laser photons) 
in the plasma is very close to the speed of light 
$v_g [= c (1-{\omega_{pe}}^2/{\omega}^2)^{1/2}]$.
In contrast to the well-known quantum coherence that can make 
particles cohere in low energy regime and high density conditions 
leading to quantum degeneracy that organizing atom condensation 
(e.g. \cite{Cohen-Tannoudji}), the relativistic coherence was understood 
to arise because when matter is forced to move in relativistic 
regime such as under LWFA, electrons are forced to accelerate in 
the wakefield potential, in which the nonlinear dynamics of the 
electron motion manifests most acutely where the potential is lowest
(after electron charge e is multiplied) by the mechanism called 
the wave-steepening. (An artistic capture of this concept was 
immortalized by Hokusai in his famous depiction of nonrelativistic
(water wave) wave-steepening\cite{Tajima2011}. This steeping would 
have led to the wave breaking in a plasma when the plasma attains 
the wave-breaking field at the Tajima-Dawson field $E_0 = m_e\omega_{pe}c/e$.
In the relativistic regime, however, we recognized due to the 
relativistic coherence arising in LWFA the wave-breaking is largely 
suppressed, because electrons cannot easily surpass the leading laser 
pulse that is speeding relativistically in the plasma with $v_g$. 
As a result, the steepening of the LWFA now avoids the wave-breaking 
and rather it leads to a cusp formation, a nonlinear singularity, 
that shows a robust wake structure (see Fig.\ref{Fig1}). 
As a result of this 
the comb-like structure with the above cusp singularity (see e.g. 
\cite{Pirozhkov2012} have exploited such singularity in emitting 
coherent radiation).
The wake wave moves with the phase velocity $v_{ph}$
which is equal to the group velocity of the photon $v_g$. 
A model equation for this forced oscillation driven by the lasers 
pondermotive potential $\Phi$ (the nonlinear $v \times B$ force on the 
right hand side) may be obtained \cite{Esarey} as 
\beq\label{eq2}
\frac{\partial^2\Phi}{\partial\zeta^2} + k^2_{pe}\Phi = 
k^2_{pe} m_ec^2 \frac{a^2(r,\zeta)}{2},
\eeq
which leads to the formation of the wakefield structure shown by
\beq\label{eq3}
E_{\omega} = \frac{\sqrt{\pi}}{4}a^2_0E_0k_{pe}\sigma_z
\exp\left(-\frac{k^2_{pe}\sigma^2_z}{4}\right),
\eeq
in a cylindrical symmetric case in $r$, $z$, the wakefield may be characterized.
Here $\zeta = z-ct$, $a=\frac{eE_l}{m_e\omega c}$, $E_l$ is the laser electric
field, $a_0$ is the maximum of $a$, $k_{pe}=\omega_{pe}/c$
and $\sigma_z$ is the laser pulse length.
In the longitudinal direction, the above mentioned 
strongly cusp-shaped robust wakefield structure develops. 
On the other hand, the transverse direction, the electric potential 
shows the betatron structure that is reminiscent of the conventional 
accelerators transverse restoring farce (see e.g. \cite{Chao-Tigner}). 
The restoring strength $K$ of betatron oscillation is given by
the radial force $F_r$
\beq\label{eq4}
K^2 = \frac{F_r}{m_e c^2r}=2\sqrt{\pi}\frac{a^2_0k_p\sigma_z}{{r_L}^2}
\exp\left(-\frac{k^2_{pe}\sigma^2_z}{4}\right)\sin k_{pe}\zeta,
\eeq
which characterizes the transverse oscillation of betatron motions 
in the $x$-direction modeled in the following equation
\beq\label{eq5}
\frac{d^2x}{dt^2}+\left(\frac{\omega_{pe}}{\gamma}\frac{E_z}{E_0}
+\tau_Rc^2K^2\right)\frac{dx}{dt}+\frac{c^2K^2}{\gamma}x=0,
\eeq
where $r_L$ is the laser width, $\gamma$ is the Lorentz factor 
of the electron, and $\sigma_z$ the laser pulse length,
$\tau_R$ is the radiation time $\frac{2e^2}{3m_e c^3}$.
Note that in this in addition to betatron restoring force, the third term, 
the second friction term also arises from accelerating effect as well as 
synchrotron radiation emission\cite{Schroeder, Nakajima2011}. 
The synchrotron radiation damping is believed to become significant only 
in very high energies (such as 10s of TeV in a typical laser acceleration 
experimental conditions). In extremely high energy regime, however, 
this betatron damping causes both the damping of this oscillatory motion 
that helps reduce the emittance of the beam in the transverse direction 
and simultaneously reduces the accelerating gradient in the longitudinal 
direction\cite{Schroeder, Nakajima2011}. The greater the transverse oscillation
amplitude $x$ is, the more significant this effect is.  
Thus electrons with greater transverse oscillation (with greater $x$) 
tends to decelerate out of the accelerating wakefield bucket, leaving only 
electrons with small betatron oscillations in the bucket\cite{Khalid2012}.  
This leads to the ever shrinking transverse emittance that are kept in 
the bucket being accelerated in extreme high energies. This opens a 
possibility that we may be able to accelerate small class of electrons 
that remain in the bucket with extremely small emittance can gain extreme 
high energies in the regime where the betatron oscillation cases 
significant synchrotron radiation damping and thus cooling. 
This may be an advantage to aim for extreme energies as considered in 
Tajima et al.\cite{PTP2011}.

\subsection{Low-density Operation of LWFA}
A number of pioneering experiments
\cite{Nakajima95,Madena95,Faure,Mangle,Leemans2006} have demonstrated 
its proof-of-principle scientifically. Its application to a future collider
has been considered\cite{Xie1997,Esarey}. 
The above proof-of-principle experiments as well as the typical collider 
design parameters have focused on the plasma density regime of 
$10^{17}-10^{18}$ cm${}^{-3}$ so far. 
This has risen chiefly from the available laser technology at that time 
and secondly from the relatively short length of acceleration required 
(and thus less demanding condition on the laser guiding over the 
acceleration distance ). The fundamental relativistic dynamics of LWFA 
determines the amplitude of the wakefield and the group velocity of the 
laser. From these all the LWFA based collider scalings as of function of 
the plasma density may be derived\cite{Nakajima2011} based on the scaling 
of the energy gain per stage $\Delta\epsilon = 2m_e c^2(n_c/n_e)$ and 
the dephasing length $L_{dp} = \lambda_0(n_c/n_e)^{3/2}$\cite{TT-Dawson},
where $n_c$ is the critical density of the laser $\omega_{pe} = \omega_0$,
$\lambda_0$ the laser wavelength, and $n_e$ is the electron density of 
the plasma.
We now list the major relationships of LWFA parameters as a function of 
the plasma density. Based on this scaling, we also check the scalings 
of multi-stage acceleration and its implications to a collider design.
The dephasing length is given by
\beq\label{eq6}
k_{pe}L_{pd} \sim \frac{8}{\sqrt{\pi}a^2_0k_{pe}\sigma_z}\frac{\omega^2_0}{\omega^2_{pe}}\exp\left(\frac{k^2_{pe}\sigma^2_z}{2}\right)
\sim\frac{7.4n_c}{a^2_0}\frac{n_c}{n_e},
\eeq
the stage length $L_{stage}$
\beq\label{eq7}
L_{stage} \sim \frac{\lambda_{pe}}{4}\frac{n_c}{n_e}
=\frac{\lambda_0}{4}\left(\frac{n_c}{n_e}\right)^{3/2},
\eeq
where $\lambda_{pe}$ is the plasma wave length $2\pi/k_{pe}$,
and the energy gain per stage $W_{stage}$ is
\beq\label{eq8}
W_{stage} \sim E_zL_{stage} = 
\frac{\pi m_e c^2}{2}\frac{E_z}{E_0}\frac{n_c}{n_e},
\eeq
the number of stages to reach the final energy of the beam $E_b$
with the Lorentz factor $\gamma_f$
\beq\label{eq9}
N_{stage} = \frac{Eb}{W_{stage}} \sim 
\frac{2\gamma_f}{\pi}\left(\frac{E_z}{E_0}\right)^{-1}
\left(\frac{n_c}{n_e}\right)^{-1}.
\eeq
The laser energy per stage is then
\beq\label{eq9}
U_L \sim P \tau_l =
\frac{\sqrt{\ln2}}{16\pi}\frac{\lambda_0}{c}\left(\frac{m^2_e c^5}{e^2}\right)
a^2_0(k_{pe} r_l)^2k_{pe}\sigma_z\left(\frac{n_c}{n_e}\right)^{3},
\eeq
where $\tau_l$ is the laser pulse length, $P$ the power of the laser,
$r_l$ the laser focal transverse size, then the wall plug power $P_{wall}$
to drive this system is
\beqa\label{eq10}
P_{wall} \sim 0.28[\mbox{MW}]\frac{a^2_0k_{pe}\sigma_z}{\eta_l}
\left(\frac{1-\eta_l}{\eta_l}\right)^2
\frac{(k_{pe} r_l)^2}{(k_{pe}\sigma_{x0})^4}
\left(\frac{E_z}{E_0}\right)^{-3} \nnb\\
\times
\frac{\sigma_x\sigma_y}{(1[\mbox{nm}])^2}\left
(\frac{E_b}{1[\mbox{TeV}]}\right)^3
\left(\frac{n_e}{10^{17}[\mbox{cm}^{-3}]}\right)^{1/2},
\eeqa
where $\eta_l$ is the laser efficiency, $\sigma_x$ and $\sigma_y$
are the laser transverse sizes in the $x$ and $y$ directions.
Note here that it is imperative to stay in the quasi-liner regime of 
LWFA (rather than the bubble regime ) for a collider design in order 
to keep the beam property pristine\cite{Xie1997,Esarey}. 

Based on this collider scaling of LWFA on the plasma density, 
Nakajima et al\cite{Nakajima2011} found that although the energy of 
the laser pulse per accelerator stage (over one dephasing length $L_{dp}$ 
is inversely proportional to $n_e ^{3/2}$, the overall average laser power 
of a collider is proportional to $n_e^{1/2}$. This means that when 
we change the plasma density from $10^{17}$ to $10^{15}$ cm${}^{-3}$, 
the energy gain per stage changes from 10 GeV to 1 TeV, the dephasing 
length from 10 cm to 100m, and the necessary laser energy from a few J 
to a few kJ, while the required average laser power reduced from 10 MW 
to 1 MW for a TeV collider. In other words, even though in this low 
density operation needs a thousand times energy per stage, the average 
laser power goes down by a factor 10, which is most likely a determining 
factor of the cost of such a collider. This points to the direction 
toward the low density operation in the LWFA collider. This is one of 
the rationales that the recent science case of the IZEST 
(International Center for Zetta- and Exawatt Science and Technology) 
dedicates its laser (PETAL with $> 3$kJ per beam ) 
to the laser electron acceleration project\cite{IZET-HP}.

However, the adoption of a kJ class laser alone will not solve the 
collider challenge, as the luminosity requirement remains to be fulfilled. 
Even at the low density operation the required laser averaged power is 
the quite high MW range. There is no laser technology that can fit this 
requirement now. In collaboration between the high energy accelerator 
community and the high intensity laser community represented by 
ICFA (International Committee for Future Accelerators) and ICUIL 
(International Committee for Ultra High Intensity Lasers) they have 
identified the future technologies that could satisfy the condition 
\cite{ICFA-Newslett-2011}. Through this and other studies emerged the 
technology of fiber lasers. Currently, the project 
ICAN (International Coherent Amplification Network)\cite{ICAN-HP} 
has been established to nurture this technology toward the application 
of the laser-driven collider (and other large averaged power applications).

\subsection{Toward Ultrahigh Energies}
Tajima et al.\cite{Tajima2011} noted that utilizing this wakefield 
in low density regime and the steeply rising wakefield in 
the relativistically intense laser (the wakefield enhancing effect 
of $a_0^s$ dependence) one can make a plausible path toward even 
Petaelectron volt (PeV) acceleration. As we mentioned above, however, 
we notice that transverse betatron amplitude also damps, which leads to 
the possibility of reduced betatron oscillation in Eq.(\ref{eq5}), 
which in turn allows less synchrotron emission by the reduction of $K^2$. 
Thus, we see a path to ultrahigh energy acceleration with ever reducing 
betatron oscillation and reduced transverse emittance in 
ultrahigh energies\cite{Khalid2012, Nakajima2011}. 
Hung and Ruth considered the ultimate Compton emittance with the emission
of photons off synchrotron radiation in the betatron oscillations\cite{PRL19}.
Recently Nakajima et al.~\cite{Khalid2012} are revisiting this issue.

       If and when such ultrahigh energies may be achieved, 
we may be able to conduct research that is not necessarily bound by 
the collider approach. In \cite{PTP2011} they suggested to explore 
the vacuum texture that may arise from the quantum gravity, for example, 
by measuring the speed of gamma photon propagation speed.  
This corresponds to checking the first term of Eq.(\ref{eq0}) 
if this term is simply nonexistent or has a modification from Einstein's 
theory of Special Relativity. If such research can be conducted, 
it is no more necessary to require high luminosity of a collider. 
Low repetition or even single shot experiments may provide a tantalizing 
evidence of such a possibility. So far only astrophysical observation of 
gamma ray burst (GRBs) that traverse the entire length of the Universe 
since its burst in the primordial times possibly showing a differential 
arrival times of different energy gamma photons from such an event. 
The 10s second differential arrival times at GeV gamma photons of GRB 
would correspond to a laboratory detection of arrival differentiation 
in fs over the distance of km for PeV gammas. 

\begin{figure}
\begin{center}
\includegraphics[width=1.0\linewidth]{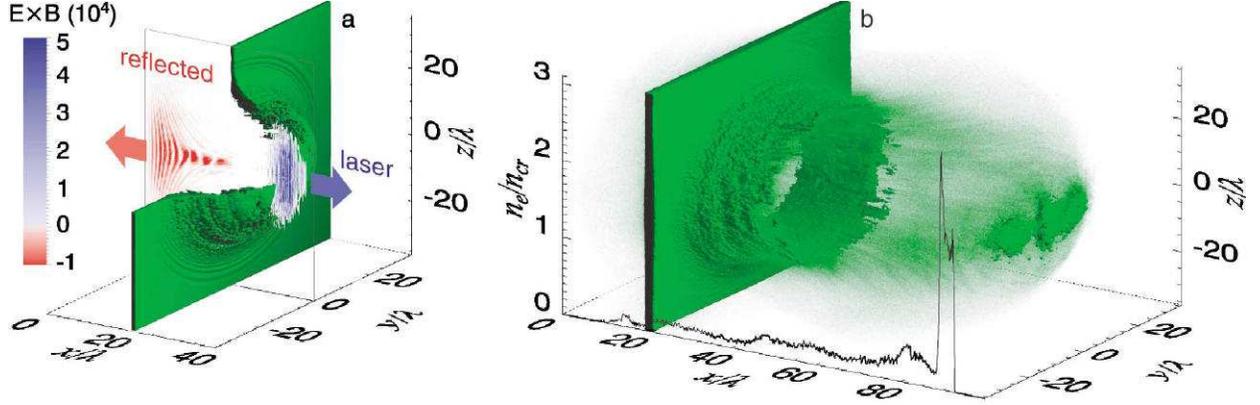}
\caption{
When the laser intensity becomes sufficiently high (exceeding on the order 
of $10^{23}$W/cm${}^2$), not only electrons become relativistic in the fields 
of laser, but also ions begin to behave relativistically, which is sometimes 
call the ultra-relativistic regime of laser intensity. In this example of 
thin high density foil laser interaction, laser pulse drives electrons ahead 
of the laser pulse by the strong light pressure, which set up a strong 
accelerating fields (on the order of TeV/m). Unlike less intense laser regime 
where ions remain non-relativistic, in this regime ions too are 
relativistically accelerated from the beginning so that ions can following 
the accelerated electrons that are strongly driven by laser, forming the 
phase-stable accelerating bucket structure, thus forming nice property of 
ion beam.  Again, the key is the relativistic coherence in the strong enough 
laser intensity\cite{Esirkepov2004}.
}
\label{Fig2}
\end{center}
\end{figure}

Ashour-Abdalla et al.\cite{Ashour-Abdalla1981} have explored 
the possibility that ultrareltivistically intense electromagnetic pulse 
far beyond the one we considered in the above for the LWFA may be capable 
of accelerating not only electrons but also positively charged particles 
and applied their theory to pulsar acceleration in which the pulsar 
wind may sustain ultrarelativistic acceleration in which the pondermotive 
potential makes what is called the snowplow acceleration. 
In the present day modern ultraintense laser development, now it appears 
eminently possible to begin to see experiments in which such a possibility 
may soon arise. In the laser intensity on the order of $10^{23}$W/cm${}^2$
the ion dynamics approaches relativistic so that ions too behave similarly 
to that of electrons. This regime may be now called the ultra-relativistic 
regime.  See Fig.\ref{Fig2}.  
Unlike the earliest experiments\cite{Snavely2000}, 
in the ultrarelativistic cases ions can be in a stable accelerating 
phase with the electron sheath that is driven by the pondermotive force 
of the laser. In this case ions can be continually accelerated. 
Esirkepov et al.\cite{Esirkepov2002} have considered such an experimental 
possibility in which an intense ($I > 10^{23}$W/cm${}^2$) can drive ions 
by driving electrons by the following equation:
\beq\label{eq11}
\frac{dp}{dt} = \frac{E^2_l [t-x(t)/c] }{2\pi n_e l} 
|\rho(\omega^{'})|^2
\frac{\sqrt{m^2_ic^2 + p^2}-p}{\sqrt{m^2_ic^2 + p^2}+p},
\eeq
where $m_i$ is the mass of the ions, $p$ their momentum, 
$d$ the length of the pulse, and $|\rho(\omega^{'})|^2$ the
reflected power fraction of the laser at the frequency of $\omega^{'}$.

 Here once again the pondermotive potential force is at play on the 
right hand side.  More recently Zheng et al.\cite{Zheng2012} showed 
that a large energy laser (on the order of 100kJ class may be capable 
of accelerating ions to sub TeV over mere distance of 10s cm). 

\subsection{Colliding Lasers for Ultrahigh Intensity}
 As mentioned above, the relativistic dynamics of the LWFA brings 
in the coherent robust wake structure that contains the cusp singularity 
of the electron density in the plasma. Since this cusp density structure 
is moving near speed of light, this high density electron density structure
may be regarded as a flying mirror that Einstein wished to use his 
Gedanken experiment of relativity. Bulanov et al.\cite{Bulanov2003} have 
introduced the relativistic flying mirror based on LWFA. 
(Recently Mourou et al.\cite{Mourou2012} have suggested plasma compressor 
of laser pulses in Exawatt and Zettawatt regimes where no longer solid state 
grating compressor is plausible, because the intense and fluence of such laser 
would burn the solid grating compressor.  Thus this plasma mirror or 
Raman plasma compressor has a function similar, to but still distinct 
in detail from this relativistic flying mirror). They showed that LWFA 
cusp density may have not only favorable longitudinal structure mentioned 
above, but also carries a parabolic curved geometry suitable for focusing 
the reflected photons off the flying mirror. In this case this flying 
mirror not only acts as the longitudinal compressor of the counterstreaming 
laser pulse that hits the wake mirror, but it has a capability to focus 
transversally.  Because of these two features, Bulanov et al. showed 
that the potential of highest intensity of laser focus may be achieved 
at the intensity of 
\beq\label{eq12}
\frac{I_{sf}}{I_{s}} \sim 32\left(\frac{\omega_d}{\omega_s}\right)^2
\left(\frac{D_s}{\lambda_s}\right)^2\gamma^3_{ph},
\eeq
where $I_{sf}$ is the intensity in the focal spot of the source
laser pulse, $I_s$ the source laser intensity, $\omega_d$ and $\omega_s$
are the frequencies of the driver and source lasers, $D_s$ is the diameter
of the flying mirror, $\lambda_s$ the wavelength of the source laser,
$\gamma_{ph}$ the Lorentz factor of the wakefield (flying mirror).
This shows that the possible intensity from the relativistic flying mirror 
may exceed that of the Schwinger field (Schwinger, 1951) beyond 
$10^{29}$W/cm${}^2$, using the exiting PW class lasers. 
A series of proof-of-principle experiments have been conducted to the 
longitudinal compression so far\cite{Kando, Pirozhkov}. 
The above is plasma-mediated two colliding lasers, one is LWFA flying 
mirror causing laser pulse, while the other is reflected by the flying 
mirror to be focused to ultrahigh intensity. 

\section{Exploring the Texture of Vacuum}
\subsection{Non-luminosity Paradigm}
 In the high energy collider approach in order to keep the number of 
events per unit time $N=\sigma {\cal L}$ (for the purpose of unifying a new 
discovery of, say, a new particle), where $\sigma$ is the cross-section 
of such an event, the luminosity ${\cal L}$ in Eq.(\ref{eq1})
has to increase to the square of 
the energy $\epsilon$, as $\sigma \propto \epsilon^2$ according to 
the Heisenberg uncertainty principle. The average power of the beam 
$P_b = f(N_1+N_2)\epsilon$, therefore, has to increase faster than the 
first power of energy $\epsilon$. The exact exponent depends upon the precise 
design of a collider. As an illustration, if we keep $A$ and $f$ constant,
and we increase $N_1=N_2 \propto \epsilon$, we have $P_b \propto \epsilon^2$. 
If we keep $A$ and $N_i$ constant and we increase $f \propto \epsilon^2$,
$P_b \propto \epsilon^3$. This shows that the development of a high energy 
collider is a tough challenge not only in its energy frontier but in its 
high luminosity aspect.

  Here we suggest a new paradigm of the non-collider approaches. 
There is a class of important contemporary physics questions that may 
be explored if we can extend the energy frontier significantly, even if 
we have insignificant or little luminosity. By removing the condition 
of the increasing luminosity ${\cal L}$ as an increasing function of 
energy $\epsilon$, we can make the high energy accelerator much 
easier (and cheaper) and thus the realization time horizon much closer. 

\subsection{Testing the Relativity in Extreme Energies}
The examination of the Lorentz invariance in the ultrahigh energy regime 
is an example without luminosity demand~\cite{PTP2011}. There have been some 
\cite{Sato,Sidharth,Ellis} that the Lorentz invariance may be broken in 
the very high energy regime, in particular toward the Planck energy. 
A possibility might be that in such high energies the vacuum becomes 
bubbly or lumpy due to quantum gravity fluctuations. If so, 
electromagnetic waves of extreme short wavelengths (or $\gamma$-photons 
with extreme high energies) may begin to feel this bumpiness, possibly 
leading to a change (say, reduction) of the observed speed of light. 
Even though it is conceivably not possible to reach for the Planck energy, 
some tell-tale tendency might appear in lower energies. In fact 
astrophysicists have been studying the delayed arrival of $\gamma$-photons 
from primordial GRB (gamma ray bursts) as a function of their energy. 
So far \cite{Ellis,Abdo} these GRB studies seem to show that the higher 
the energy of $\gamma$-photon of a particular GRB is, the greater the 
delay of arrival is by 10 to $10^2$s. However, the astrophysical 
observations cannot resolve the question if this delay is in fact 
originating from the high energy texture of vacuum (i.e. indicating a 
possible violation of the Lorentz invariance) or it shows the mechanism 
of creating (or emitting ) higher energy $\gamma$-photons. A control 
terrestrial laboratory experiment in ultrahigh energies may resolve this 
question. By replacing the distance to the edge of the Universe 
$10^{28}$cm by a km ($10^5$cm) and the energy of photons in GeV by PeV, 
the time resolution of the delayed arrival down to 1~fs from 10~s can be 
comparable in determining the above issue. (If the $\gamma$-photon energy 
is 10~TeV, we need the time resolution of 10~as).  Such an attempt may be 
regarded as the exploration of the nature manifested through the first 
term of vacuum susceptibility in Eq.(\ref{eq0}). 

Tajima et al.~\cite{PTP2011} have suggested to accelerate electrons 
toward PeV by LWFA, utilizing a MJ laser 
(such as the LMJ laser~\cite{LeGarrec}). The conventional 
RF acceleration may have a difficulty to reach PeV energies, as the 
synchrotron radiation power loss begins to become formidable beyond 
30~TeV~\cite{Ostermayr}.

 In addition Altschul~\cite{Altschul} suggests that the spectrum of 
synchrotron radiation can begin to deviate from the known power law 
as a function to a steeper energy loss in the higher energy 
(beyond around 30~TeV)
\beq
P(\gamma) = P_0(\gamma)(1+\gamma^2\delta),
\eeq
where $\delta$ is believed to be less than $10^{-15}$
(eliminations by astrophysical observations).
Since the LWFA makes betatron oscillations of electrons in the wake field 
emitting synchrotron radiation\cite{Dino}, the measurement of 
the synchrotron spectrum beyond 30TeV may also be of interest.

\section{Zeptosecond Streaking and a Study of the Structure of Vacuum}

  Another characteristics of the departure of the collider paradigm 
we are embarking on in addition to the previous section is the potential 
to trace the examination of relevant physics 
in the time development of the interested phenomenon. 
In the collider paradigm one studies the spectroscopy of particle 
interaction and creation, {\it i.e.}, the energy dependence of the 
particle dynamics. If the probe preserves both the amplitude and the 
phase of the recorded signal, the spectrum in $\omega$ is just the 
Fourier transform of the time series in $t$ and both quantities are 
related and equivalent. However, in most collision events, the phase 
information is not retained so that the obtained spectrum from a collider 
cannot tell the temporal history. The introduction of coherent photons 
(laser) as a probe gives us the possibility to retrieve the phase 
information. This is particularly powerful when our target of probe is 
vacuum itself, as noise in largely suppressed due to the lack of 
interaction with charged matter.

\subsection{Zeptosecond Streaking of Vacuum}
  In atomic physics based on laser, both the energy-resolving approach 
(spectroscopy) and the time resolving approach do exist, though the former 
developed first. In the latter the temporally evolution of atomic dynamics 
has been studied in fs (the laser period scale) and nowadays even in as 
(much less than the laser period)\cite{Goulielmakis2008}, in which the laser 
pulse streaks the atomic electron to map its temporal dynamics 
(the attosecond streaking). With the aid of laser coherence the electron 
dynamics in an atom is embedded in the electron momentum which preserves 
the phase information. The modulated momentum of the electron is now 
mapped again into the time domain by the time-of-flight measurement of 
the electron. One can tell which time the electron interacted with photon 
through the phase information.

Real electron-positron pairs are created out of vacuum at
the Schwinger critical field\cite{Schwinger} 
\beq\label{eq14}
E_{u.c.} = \frac{m^2 c^3}{e\hbar}=1.3 \times 10^{18} \mbox{V/m}
= 4\times 10^{29} \mbox{W/cm}{}^2.
\label{eq_Es}
\eeq
This pair creation mechanism should be distinguished from that of
the perturbative or point-like interaction. An external electric 
field $E$ causes the pair creation via the tunneling effect in 
the Dirac sea of electrons in vacuum. The rate of the pair creation 
in the external electric field $E$ is proportional to the following 
tunneling factor $\exp\{-\pi E_{u.c.}/E\}$, which corresponds to the 
non-perturbative description of the phenomenon.
Due to the strong exponential suppression, we are not able to expect 
the phenomenon with reasonable event rate in laboratories, 
unless $E$ exceeds $E_{u.c.}$. However, if high-intensity lasers are used 
combined with energetic photon or charged particle, we can effectively 
lower the tunneling probability for electron-positron pair to be created 
in the non-perturbative regime from the laser induced 
vacuum~\cite{Narozhny}. The production rate $R_{e^+e^-}$is known as
\begin{eqnarray}\label{eq15}
R_{e^+e^-} = \frac{e^2 {E_l}^2}{4\pi^3}
\exp \{ 
-\frac{8}{3} \frac{E_{u.c.}}{E_l} \frac{m_e c^2}{\hbar\omega_{\gamma}}
\},
\end{eqnarray}
where $E_l$ is the electric field of laser and $\hbar\omega_{\gamma}$ is the 
incident $\gamma$-ray energy. If an electron is created and incident
in this process, prompt 
radiations of gammas off the electron immediately create successive 
pairs and gamma radiations, resulting in an electromagnetic shower 
inside the laser field\cite{Ruhl}. 
A study of such a radiation mechanism under 
an extremely high-acceleration fields is of contemporary interest as
it is also related to the horizon radiation\cite{53a}. 
It is of interest of nonlinear QED, but it has further implications 
as we discuss later.

  It is possible to concoct two counterstreaming intense lasers 
(laser1 and laser2) in vacuum to investigate the property of vacuum 
in ultrahigh intensity electromagnetic fields if we impose another 
high energy gamma photons co-propagating with one of the laser 
pulse\cite{Ipp2011,Tajima2012}.
As compared with laser photons (energy of eV), the high-energy gamma
photons carrying energies of, say GeV (or even 10GeV) have so lopsided
momenta compared with the one of the lasers that is counterstreaming
against the gamma photons. What we call the Nikishov frame is
transformed toward the laser pulse with the gamma factor
of $\gamma_N = \hbar w_{\gamma} / m_e c^2$\cite{Tajima2012}.
On this frame the laser pulse that is counterstreaming
against the gamma photons is Lorentz contracted in its pulse length by
$\gamma_N$, while the electric field of the laser pulse is
Lorentz transformed to an enhanced value of $E^{'}_l = E_l\gamma_N$.
This process is motivated by the Nikishov-Ritus expression of the
probability to break down the vacuum to produce electron-positron 
pairs~\cite{Nikishov-Ritsu}, as discussed in Eq.(\ref{eq15}), 
and may be regarded as an interpretation of Eq.(\ref{eq15}).
The Schwinger invariant $E^2-B^2$ that is zero for a single plane
wave laser pulse now doe not vanish any more because of two
counterstreaming laser pulses.
When we impose that gamma photons are embedded in the co-propagating
laser pulse and these two encounter simultaneous at one particular
spacetime, they now see this compressed and enhanced electric field of
the counterpropagating laser pulse. Thus by choosing sufficiently
intense laser pulse counterstreaming against the gamma photons.
This setup allows us to access and approach the Schwinger field in
vacuum with existing laser intensity. Furthermore, more importantly,
because of the counterstreaming laser pulse length may be regarded Lorentz
compressed from the value of $\tau^{'}_{L1} = \tau_{L1}/\gamma_N$.
The single oscillatory period of  few fs ($\tau_{L1}$) is now compressed to
a few as (if the gamma photon energy is GeV (is 100 of a few zs if it
is 10GeV). Thus the fractional phase within a single period of the
laser1 is now in the regime of zs on the Nikishov frame. Thus if and when
the laser1 field exceeds (or becomes close to) the Schwinger value,
the electron-positron pair production results and these particle are
now driven by the combination of the co-propagating and counterstreaming
laser pulses laser 2 and laser 1.  This is very much akin to the situation
of atomic attosecond streaking\cite{Goulielmakis2008}.
In these laser fields the Nikishov laser field
(in the normalized vector potential of the laser1) now takes the value of
what we call the Schwinger-Nikishov amplitude
$a^{SN}_0 = eE^{SN}/m\omega_0c$ as
\beq\label{eq13}
a^{SN}_0 = \frac{m_ec^2}{\hbar\omega_l}\frac{m_ec^2}{\hbar\omega_{\gamma}},
\eeq
where $\omega_r$ is the frequency of the gamma photon.
Electrons (and positrons) thus created would be accelerated by the imposed
EM field of these two lasers, producing streaking patterns of their momenta.

  The application of the attosecond streaking to the streaking of vacuum 
has been considered by \cite{Ipp2011} and others. 
It has been suggested that the injection of high energy $\gamma$-beam 
(GeV or more ) synchronized with two counter streaming large amplitude 
laser pulses in order to see a streaking pattern of electrons 
(and positrons) that are generated by the pair creation out of vacuum 
(the process of Nikishov-Ritus-Narozhny). 
As Mourou and Tajima\cite{54a} have suggested the Pulse-Duration Intensity
Conjecture, such an attempt toward ultrafast pulses may be regarded as
an example of realization of even shorter pulses with even more intense
laser pulses.  Because the laser pulse that is 
counter-propagating against the $\gamma$-photon is Lorentz compressed 
in the pulse length and boosted in the Nikishov frame. 
The detailed streaking pattern should reflect the 
electron (and positron) dynamics, including their emergence out of vacuum 
reflecting the Schwinger process in time. 

If we vary the range of the probe $\gamma$ energy from MeV to even 
beyond TeV scale, examining the quark pair creation in high-intensity 
laser fields comes into view. In this case a curious issue arises as to 
the mass of quarks. Normally quarks are confined by essentially 
chromo-fields in a baryon or a meson as the constituent of hadrons and 
their masses are on the order of the hadronic masses. On the other hand, 
quarks have electric charges and also should be expressed by the Dirac 
equation insofar as the electric charge is concerned. The corresponding 
Schwinger critical field strength should scale as $(m_q/m_e)^2$ and $e/q$ 
where $m_q$ is quark mass and $q$ is the quark electric charge. 
In addition, if we regard the factor $m_ec^2 / E_{\gamma}$ in the exponent of 
the tunneling probability in Eq.(\ref{eq15}) as the inverse Lorentz factor of 
carriers of virtual vacuum polarization on the Nikishov frame, we may 
replace the electron mass by that of quark. Therefore, the tunneling 
probability should strongly depend on the dynamical quark mass while it 
emerges from the initial current mass in vacuum to the constituent mass 
confined in hadrons after the pair production. The suggested vacuum 
streaking method with extremely high-frequency $\gamma$-probe may provide 
a novel approach to scan the dynamical structure of such an evolution 
of mass. We may even imagine that this method might  be applied to scan 
the mass evolution in the emergence from vacuum in the pair production 
of an electron a positron. Since the normal electron mass is not 
determined by QED itself, we may also try to check the dynamical electron 
mass evolution by testing the threshold of the real pair creation as 
a function of incident $\gamma$ frequency. In addition to the standard 
view point based on the Higgs scenario to determine Fermion masses via 
the Yukawa-coupling with a Higgs field, this kind of test would add, 
at least, a novel insight, as considered by Dirac sometime ago\cite{Dirac},
from different aspect on how the mass is dynamically generated from 
massless field and also how the concept of mass, charge, 
and the background acceleration field are related.

It is also interesting to realize super-Schwinger field.
It is possible to do so, because when the Schwinger invariants vanish no
vacuum breakdown happens. As Narozhny's expression\cite{Narozhny2004} shows,
we could design the setup of lasers (and / or other EM-fields) in such a way
to reach super-Schwinger field without breakdown only to trigger the breakdown
or other relevant phenomena in a controlled fashion.

\subsection{Phase Contrast Fourier Imaging of the Laser-Induced Vacuum}
The pair creation discussed so far is related to the imaginary part 
of the dispersion relation of the laser-induced vacuum. This information 
provides the perspective to quantitatively understand, certainly, the QED 
dispersion relation~\cite{Shore} combined with the measurement of that 
of the real part. This is directly connected with the concept of the 
1st order polarization susceptibility $\chi_1(\omega)$ , as we discussed 
in Eq.(\ref{eq0}) of Introduction. 
The real part can be measured via the elastic 
photon-photon scattering process. If two photons collide 
head-on, the center of mass system (cms) energy is on the order of eV. 
In such a low cms energy the total elastic scattering cross section is 
on the order of $10^{-42}$~b\cite{KN,DT}. This is too small to detect the 
scattering process at a large scattering angle. Instead, we should be able to 
measure the change of the refractive index of vacuum induced by the 
high-intensity target laser field by crossing the probe laser field,
where we focus on only the forward scattering amplitude between 
the target and the probe laser beams. The change of the refractive index is 
proportional to the energy density of the target laser. For example, 
the amount of the change is $10^{-10}-10^{-9}$ for a 200~J pulse per 20 
fsec time duration\cite{APB-QED}. Homma et al. proposed the method: 
if a sufficiently high-intensity probe laser is available, it is not 
impossible to measure the effect on the pulse-by-pulse basis\cite{APB-QED}. 
This is the phase contrast imaging by the probe laser of the warping of 
vacuum distorted by the intense pump laser field. Once we could measure it,
it is possible to discuss the dependence on the combinations of 
polarizations between the target and probe lasers, {\it i.e.}, 
the birefringence of the laser induced vacuum. The ratio of the two 
cross-sections of the parallel and normal crossings is 4:7 in the QED
refractive index change. This ratio comes from the 
one-loop effective Lagrangian, Euler-Heisenberg Lagrangian\cite{EH,Weiscop},
expressed as
\beqa\label{eq16}
    L_{1-loop} =
    \frac{1}{360}\frac{\alpha^2}{m_e^4}
    [4(F_{\mu\nu}F^{\mu\nu})^2+7(F_{\mu\nu}\tilde{F}^{\mu\nu})^2],
\eeqa
where
$\alpha=e^2/(\hbar c)$ is the fine structure constant,
$F_{\mu\nu} = \partial A_{\mu} / \partial x^{\nu} -
                    \partial A_{\nu} / \partial x^{\mu}$ is
the antisymmetric field strength tensor and its dual tensor
$\tilde{F}^{\mu\nu} = 1/2 \varepsilon^{\mu\nu\iota\rho} F_{\iota\rho}$
with the Levi-Civita symbol $\varepsilon^{\mu\nu\iota\rho}$.
It is interesting to prove or disprove the effective Lagrangian by 
directly measuring birefringence. If a deviation from the QED prediction 
is found, it indicates contributions possibly from QCD and most likely from
other new interactions due to the extremely small cross section expected 
by the QED processes. If hidden low-mass scalar or pseudo scalar fields 
could exist in vacuum, they may contribute to the first and the second 
term of the Lagrangian, respectively, and they change the expected ratio. 
However, this way to search for something beyond the standard model is 
rather limited due to the cms energy around 1eV. This limitation on the 
accessible cms energy is resolved by the method proposed in the next section. 

\section{Vacuum Optical Parametric Effect relevant to Cosmology and Particle Physics}
Contemporary cosmology supported by astronomical observations suggests
that the energy density of the universe is occupied dominantly by
unseen stuff; Dark Matter (DM) and Dark Energy (DE),
while only 4\% is explained by the ordinary matter density.
Understanding of these dark components is one of the most challenging
and intriguing subjects.

With high-intensity lasers, we expect to be able to introduce the possibility to
understand these dark components as undiscovered fields characterized by
the generic nature; low-mass and weakly coupling to ordinary matter.
Needless to say, there are is a wide variety of theoretical speculations and
it is too premature to stick to a specific theoretical model.
However, we motivate ourselves to target on the testable predictions 
accessible by both terrestrial laboratory experiments and astronomical 
observations. In particular, here we focus on the testable models which have
solid theoretical foundations based on the concept of the
spontaneous breaking of fundamental symmetry.

In this section, after reviewing two theoretical predictions,
we review the fundamentals of the new approach with high-intensity
laser fields to test these predictions.

\subsection{Theoretical predictions based on the spontaneous breaking of
the fundamental symmetry}
As for DM, axion is one of the most theoretically well-motivated cold dark
matter candidates based on the strong $CP$ problem in quantum 
chromodynamics (QCD)\cite{AxionReview}. The low-energy phenomena seen in QCD
cannot be understood only by the perturbative part of the QCD Lagrangian. 
For instance, the typical phenomenon is represented by the decay process 
$\pi^0\rightarrow2\gamma$. Adler-Bell-Jackiew\cite{ABJ} pointed out that 
the perturbative treatment of the $\pi^0-2\gamma$ coupling via the triangle
Fermion loop accompanies a pathological divergence, which is an example how 
a quantum correction violates the symmetry preserved at the Lagrangian level.
In order to cure the anomaly, it is unavoidable to implement the 
non-perturbative aspect of the QCD Lagrangian\cite{StrongCP}.

't Hooft then proposed a concept of four-dimensional Euclidean-gauge soliton
referred to as instanton, which introduces a topological degree of freedom 
as the non-perturbative part of the QCD vacuum as follows\cite{tHooft}
\beq\label{eqInstanton}
L_{\theta}=\theta\frac{g^2_s}{32\pi^2}
\tilde{G}_{\mu\nu}^{a} {G^{\mu\nu}}^{a},
\eeq
where $\tilde{G}_{\mu\nu}^{a}$ is the strong field strength tensor 
specified with the color index $a$ and $\tilde{G}_{\mu\nu}^{a}$ is its dual 
tensor defined as $\tilde{G}_{\mu\nu}^{a} = 1/2 \varepsilon_{\mu\nu\iota\rho} 
{G^{\iota\rho}}^{a}$, $g_s$ is the coupling, and
the arbitrary phase $\theta$ forms the following topological vacuum state 
\beq\label{eqTheta}
|\theta\rangle = \sum_n e^{i\theta n} |\theta\rangle,
\eeq
where $n$ is the winding number and the instanton is associated with
the tunneling process between different winding numbers.
 
On the other hand, the phase $\theta$ is also associated with the chiral 
transformation of the quark mass term which produces the same type of 
the Lagrangian form as in Eq.(\ref{eqInstanton}). 
In the standard model of particle physics,
quark masses are created via the Higgs mechanism after the electroweak gauge 
invariance is spontaneously broken and the quark mass is expressed as 
the product between quark coupling and the vacuum expectation value $v$ 
of the minimum of the scalar Higgs potential\cite{Higgs}. The quark mass
matrix $M_q$ must be diagonalized to get the quark-mass eigenstate from 
the weak eigenstate. The mass term of the Lagrangian is expressed as
\beq\label{eqMquark}
-{\cal L}_{M_q} =  \bar{\Psi}_L M_q\Psi_R + \bar{\Psi}_RM_q^{}\Psi_L
=
\cos\theta_{EW} (\bar{\Psi}_L\Psi_R + \bar{\Psi}_R\Psi_L)
+i \sin\theta_{EW} (\bar{\Psi}_L\Psi_R - \bar{\Psi}_R\Psi_L),
\eeq
with $\theta_{EW} = arg[det M_q]$. In order to make the mass matrix real, 
the adjustment of $\theta_{EW}$ by chiral transformation 
$e^{-i\theta_{EW}\gamma^5}$ is required.

As seen in the above arguments, there are at least two independent reasons 
to introduce $\theta$ in the QCD Lagrangian. Therefore, the physically realized 
phase $\bar{\theta}$ should be expressed as 
$\bar{\theta} = \theta + \theta_{EW}$. 
There is no reason to cancel out each other {\it a priori}. 
Nevertheless, the precise measurement of the neutron dipole moment suggests 
the upper limit is $\bar{\theta} < 10^{-9}$\cite{NeutronDipole}.
If phase $\theta$ is non-zero, the Lagrangian term causes $CP$-violation.
Why the phase $\theta$ is unnaturally small, or equivalently, 
why the $CP$-symmetry is so precisely conserved in the QCD Lagrangian ? 
This is the strong CP problem.
 
As a resolution to this problem, Peccei-Quinn (PQ) symmetry is proposed
\cite{PQ}. It is the introduction of an axial global $U(1)$ 
symmetry in addition to Yukawa $U(1)$ gauge symmetry in the standard model 
before the electroweak symmetry is broken. 
If such a global chiral symmetry is introduced in advance, we can choose 
a proper $\theta$ so that the $CP$-conserving $\bar{\theta}$
is physically realized independent of the spontaneous breaking of
the gauge symmetry. When the PQ-symmetry is broken, 
a low-mass pseudoscalar boson should appear as the Nambu-Goldstone boson
which is realized by Weinberg\cite{Weinberg} and Wilczek\cite{Wilczek} 
and referred to as axion. The standard axion scenario where the PQ-symmetry
breaking occurs at the energy scale of the electroweak symmetry breaking is 
already excluded. Instead, invisible axion is the focus of the current 
experimental search. In the invisible case the PQ-symmetry breaking 
scale is arbitrarily higher than that of the electroweak symmetry breaking 
scale. In such a case the coupling to ordinary matter becomes extremely 
weak and the mass becomes lighter.

Axion cosmology suggests the following generic 
scenario\cite{AxionCosmology}.
Initially the PQ-symmetry breaking occurs at a temperature $T \sim f_{PQ}$,
where $f_{PQ}$ is the vacuum expectation value associated with
the PQ-symmetry breaking.
The symmetry breaking is described by introducing a complex scalar 
potential $V(\vec{\phi})$ carrying PQ-charge. 
At $T\sim f_{PQ}$ the potential is expressed as
\beq\label{eqScalar}
V(\vec{\phi}) = \lambda(|\vec{\phi}|^2 - f^2_{PQ}/2)^2,
\eeq
which is minimized at $\langle|\vec{\phi}|\rangle=f_{PQ}/\sqrt{2}$.
When the PQ symmetry is broken, the phase $\theta$, the argument of 
$\langle|\vec{\phi}|\rangle$, is arbitrarily chosen. 
In other words, $V(\vec{\phi})$ is independent of $\theta$. 
This degree of freedom $a \equiv \theta f_{PQ}/n$ with the winding number $n$
corresponds to the massless axion. 
As the temperature approaches $T\sim\Lambda_{QCD}$,
where $\Lambda_{QCD}$ is the energy scale at which
hadrons with finite masses emerge by confining quarks inside them,
the axion begins to gain its mass by tilting the wine-bottle potential 
so that the newly emerged minimum coincides with the $CP$-conserving 
$\bar{\theta}$ by the instanton effect included in the QCD Lagrangian.
If the axion mass and the coupling to matter is within a certain domain,
it may become dominant cold dark matter.

If the PQ-symmetry breaking occurs during the inflation of the universe,
it may leave observational effects on isocurvature 
fluctuations of the microwave background\cite{isocurvature}.
In addition, if axion exists, it may couple to photons as we will define
its coupling later. Combined with such astronomical observables, 
it is crucial for the ground based experiments to directly probe the 
invisible axion field especially in the low-mass and weakly coupling domain.

Let us move on to the introduction of a testable theoretical model
concerning Dark Energy (DE). The model is based on the spontaneous breaking
of a global conformal invariance or scale invariance a.k.a. dilaton
as follows.
The Einstein-Hilbert action in the observational conformal frame 
can be described as
\begin{equation}
L_{EH} =
\sqrt{-g_{obs}} \left( \frac{1}{2}R_{obs} - \Lambda_{obs} + {L_{obs}}_M\right),
\end{equation}
where
$R$ is the Ricci scalar, $\Lambda$ is the cosmological constant,
$L_M$ is the matter Lagrangian, and
$obs$ specifies that we are choosing the special conformal frame where
the gravitational constant $G$ included in $R$ and Fermion masses included 
in the matter Lagrangian part are seen as constant\cite{STTL}.
Dirac raised a question whether the fundamental constant $G$  
has a time-varying nature or not. If it decreases as a function of the
age of the universe $t$, we may understand why $G$ is so small compared 
with the other fundamental constants, which is known as Dirac's large 
number hypothesis\cite{DiracLargeNumber}.
Jordan then invented a concrete model\cite{Jordan} by introducing 
a scalar field $\phi$ to introduce the time-varying nature, 
which is named as Scalar-Tensor Theory (STT)\cite{STTL}. 
Furthermore, by adding the $\Lambda$ term to STT, 
we discuss a today's version of the Lagrangian named as STT$\Lambda$ as follows
\begin{equation}
L_{STT\Lambda} =
\sqrt{-g} \left( \frac{1}{2}\xi\phi^2R -
\epsilon \frac{1}{2}g^{\mu\nu}\partial_{\mu}\phi\partial_{\nu}\phi
- \Lambda + L_{M}\right),
\end{equation}
where $\phi$ is the scalar field to be a source of DE, 
and the parameters $\epsilon$ and $\xi$
are related to the well-known symbol $\omega$ in the way of
$\epsilon = Sgn(\omega) = \pm 1$ and $\xi = 1/(4|\omega|) > 0$.
This form includes time-varying gravitational constant 
$G_{eff}=(8\pi \xi \phi(t)^2)$, where $\phi$ varies as a function of time.
In the Jordan conformal frame, however, the cosmological equations allows 
only the static universe\cite{Jordan} without any expansion.
Therefore, this is not acceptable as a physical model. 
On the other hand, the Lagrangian form was rediscovered
by the effective Lagrangian deduced from higher dimensional
string theories\cite{STRING}. In this sense the Jordan frame is still useful
as a theoretical frame. An important feature of the Lagrangian is
that all terms except $\Lambda$ are dimensionless.
In other words the Lagrangian form has a nature of scale invariance.

A way to naturally introduce a dimensional constant is the
conformal transformation. If we introduce a global conformal transform by 
${g_{*}}_{\mu\nu} = \Omega^2 g_{\mu\nu}$ and choose $\Omega = \xi\phi^2$
as a special case, the Lagrangian in the Jordan frame can be transformed
into the following form\cite{STTL}
\begin{equation}
L_{*} = \sqrt{-g_{*}} \left( \frac{1}{2}R_{*} - Sgn(\zeta^{-2})
\frac{1}{2}{g_{*}}^{\mu\nu}\partial_{\mu}\varphi\partial_{\nu}\varphi
- \Lambda_{*}(\varphi) + {L_{*}}_{M}\right),
\end{equation}
where $\varphi$ is related to $\phi$ in the way of
$\phi = \xi^{-1/2}e^{\zeta\varphi}$ with
$\zeta^{-2} = 6 + \epsilon\xi^{-1} = 6 + 4\omega$, and 
$\Lambda(\varphi) \equiv Ae^{-4\zeta\varphi}$.
The first term coincides with that of the physical frame where
we lose the nature of varying $G$. Instead, it obtains the nature that
the corresponding $\Lambda_*$ term can be expressed as a function $\varphi$.
Surprisingly, solutions in cosmological equations with this Lagrangian
can provide a global decaying behavior 
$\Lambda_{*} \sim {t_{*}}^{-2}$\cite{STTL}.
We should keep in mind that the choice of the time dependence of $\phi$
in STT$\Lambda$ was arbitrary, unless the conformal transformation is
introduced, which induces a scale invariance breaking via
the choice of the conformal frame, that is, emergence of the dimensional 
constant. This should be distinguished from other types of scalar models 
where the time variation of $\phi$ is introduced for the phenomenological
adjustment to explain astronomical observations without profound principles.

The coupling of $\varphi$ to matter in $ {L_{*}}_{M}$ is a crucial crossroad.
If the coupling is not allowed as introduced in the Brans-Dicke 
model\cite{BD}, the most well-known STT which respects Weak Equivalence 
Principle (WEP), the physical Fermion mass $m_{*}$ is not constant any more 
which behaves like $m_{*} \sim t^{-1/2}_{*}$\cite{STTL}. 
This is not acceptable, since any physical observations are based on the 
frame where mass is constant by which one can define a tick of clock 
common to both near and far observational points. 
On the other hand, if we allow the Yukawa-type coupling to matter, 
remarkably, we can recover constant $m_{*}$\cite{STTL}.
Therefore, we can completely reproduce the features in the observational frame
by the conformal transformation of STT$\Lambda$ with WEP violating
nature, which gives us a natural interpretation of DE without the fine-tuning.
The age of the universe is known to be $t_{obs} \sim 10^{60}$\cite{STTL} and
the observed dark energy is $\Lambda_{obs} \sim 10^{-120}$
in the reduced Planckian unit with 
$c=\hbar=M_P=(8\pi G)^{-1/2}=1$\cite{STTL}.
If $\Lambda_{obs}$ could be a decaying constant
as a function of the age of the universe such as 
$\Lambda_{obs} \sim {t_{obs}}^{-2}$, the smallness would be readily 
understandable. It is simply because our universe is old.

Although this coupling violates WEP, detailed analysis shows the violation
takes place only via quantum process including anomaly explained 
above\cite{STTL} and never happens at the classical limit\cite{STTL}. 
Therefore, an important experimental clue is to search for WEP violating 
processes through the $\varphi$-matter coupling via quantum anomaly coupling.

The scalar field couples with other microscopic fields as weakly as
gravity. It also shows  no immunity against acquiring a nonzero mass  due
to the self-energy. A simple one-loop diagram in which the light quarks and
leptons with a typical mass $m_{\rm q}\sim {\rm MeV}$ couple to the
scalar field with the gravitational coupling with the strength $\sim
\MP^{-1}$ produces the mass $m_\phi$ given by
\beq
m_\phi^2 \sim \frac{m_{\rm q}^2 M_{\rm ssb}^2}{M_{\rm P}^2}\sim (10^{-9}{\rm eV})^2,
\label{mass1}
\eeq
where we have included the effective cutoff coming from the
super-symmetry-breaking mass-scale $ M_{\rm ssb} \sim {\rm TeV}$,
by allowing a latitude of  the few orders of magnitude\cite{STTL}.  
The force-range turns out to be $m_\phi^{-1}$ which has a macroscopic size
$\sim 100~{\rm m}$\cite{STTL,nat}.
We will discuss its implication of this in our suggested measuring
technique of such fields below.

\subsection{Dynamics of dark fields coupling to two photons}
For the testable theoretical predictions above, 
we suggest searching for these dark field components 
by utilizing four-wave mixing process of high-intensity laser fields.
The coupling of dark fields to two photons is generally characterized 
by the effective interaction Lagrangians with the
low-mass scalar $\phi$ or pseudoscalar $\sigma$ fields, respectively. 
The interaction Lagrangian terms can be parametrized as
\beqa\label{eqScalar}
-L_{\phi}=g M^{-1} \frac{1}{4}F_{\mu\nu}F^{\mu\nu} \phi,
\eeqa
\beqa\label{eqPscalar}
-L_{\sigma}=g M^{-1}
\frac{1}{4}F_{\mu\nu}\tilde{F}^{\mu\nu} \sigma,
\eeqa
where $M$ has the dimension of mass and $g$ is a dimensionless constant. 
If $M$ is within $10^{11} - 10^{16}$ GeV and the mass of this dark fields
is within meV - $\mu$eV range, the dark field may become a cold dark matter 
candidate based on the invisible axion scenario \cite{AxionCDM}. 
If $M$ corresponds to the Planck mass $M_P \sim 10^{18}$~GeV, 
the interaction is as weak as that of gravity. Furthermore,
if the mass of the dark field is around the neV range, the field
may have a great relevance to Dark Energy\cite{DEptp}.
Depending on the allowed spin combinations of two-photon coupling with 
dark fields, we can argue whether the dark fields are scalar-type or 
pseudoscalar-type in general. 

\subsection{Past experimental approaches in laboratories and beyond}
The-state-of-the-art methods to search for light dark matter in terrestrial 
laboratories by utilizing the two photon coupling are the Light 
Shining through a Wall (LSW) and the solar axion search. In LSW a laser 
pulse together with a static magnetic field produces axion and the axion 
penetrates through an opaque wall thanks to the weak coupling with matter, 
and it then regenerates a photon via the coupling with the static magnetic 
field located over the wall. The solar axion search is similar to LSW, 
but different for the production location. In the Sun two incoherent 
photons may produce axions and the long-lived axions penetrate the Sun and 
the atmosphere of the earth, and they regenerate photons through the coupling 
to a prepared static magnetic field on the earth. LSW and the solar axion 
search as well as astrophysical considerations provide constraints on 
the allowed mass-range of axions and the coupling to matter.
The current upper limit on the coupling of axion-like particle to photons is
around $10^{-11}$GeV${}^{-1}$ for mass range below 1eV.

In the past experiments probing a deviation from the Newtonian potential form,
massive and huge bodies were used as test probes~\cite{WEPV}.
However, when they measure the gravitational effects at a short distance, 
they must suffer from the background physical process like the Coulomb force. 
In contrast if we use photon-photon scattering as a probe for 
such a new kind of force, the experiment would be free from the background 
physical process, since the total photon-photon elastic cross section 
in the optical energy is as small as $10^{-42}$~b~\cite{DT}.
However, a huge challenge remains, because the huge and massive probes 
are needed to have a sensitivity to the gravitational 
coupling strength.  Nevertheless, if we overcome this drawback to 
use photons as the test probe, the method opens up a new window to 
probing of weakly coupling and low-mass fields (finite long range force). 

Homma et al. proposed to explore the new domain far beyond the existing 
limits on the mass-coupling parameter space ($m$, $g/M$) by 
searching for resonantly produced low-mass particle with 
high-intensity laser fields based on the four-wave mixing process 
in the vacuum\cite{DEapb}. 
The dominant mechanisms of the enhancement on the sensitivity to $(m, g/M)$
in the proposed method are two-fold. The first is the resonant production of 
low-mass fields. As long as the field has a finite mass and coupling 
to matter, we can directly produce low-mass fields as resonance states 
such as the Higgs particle production in the high energy collider.
The production cross section is in principle free from the constraints
by the weak coupling, if the center of mass system energy $E_{cms}$ of two
colliding photons is exactly adjusted to the resonance energy.
The second one is the induction of the decay of the produced particle by 
supplying a high-intensity coherent field in the background
simultaneously with the production of the particle.
This induced process hugely enhances the signal and suppresses the noise.

In the following subsections we discuss how to realize such an extremely
low-energy photon-photon collision system and the sensitivity to the 
coupling as weak as gravitational coupling for the mass range well below 
optical frequency 1~eV. This method is actually
quite similar to the four-wave mixing process in quantum optics.

\subsection{Quasi-Parallel System of photon-photon collisions}
When we are interested in extremely low-mass ranges below 1eV,
we need to reduce the cms energy of photon-photon collisions
compared to the incident photon energy assuming optical laser fields.
As long as the resonance allows the decay into only two photons,
the scattering process looks like elastic scattering
even if a low-mass field is exchanged via the resonance state in cms.
Thus there is no frequency shift in the final state in cms.
However, if we boost this system to the direction
perpendicular to the colliding axis, the frequency shift takes place along
that boost axis, as illustrated in Fig.\ref{Fig3}. 
In the forward direction on the boost axis
we expect a frequency up-shift close to the double the incident 
frequency, while a zero frequency photon must be emitted to the backward 
direction due to the energy-momentum conservation, independent of
dynamics of the exchanged field.
We refer to this boosted system as the quasi-parallel system (qps).
This may be interpreted as if the second harmonic photon is
generated from the nonlinear vacuum response.
This may be an interesting analogy to second harmonic generation 
due to the nonlinear response of a crystal with a
laser injection which was pioneered by Franken et al.~\cite{Franken}.
Inversely, if we realize the qps in the laboratory frame,
the corresponding cms energy can be very much lowered.
The cms energy with variables in qps can be defined as
\begin{equation}\label{eq_1}
E_{cms} \sim 2\vartheta \omega,
\end{equation}
where $\vartheta$ is defined as half the incident angle between two incoming
photons with $\vartheta\ll1$ and
$\omega$ is the beam energy in unit of $\hbar = c = 1$.
This relation indicates that we have two experimental handles to
adjust $E_{cms}$. If we take the head-on collision geometry,
we have to introduce very long wavelengths for the incident photons.
However, it is not too difficult
to introduce the very small incident angle. In such a case $E_{cms}$
can be lowered by keeping $\omega$ constant.
We also know that the cross section $\sigma_{qed}$ of photo-photon scattering
in the QED process in qps is largely suppressed due to
the fourth power dependence on the incident angle which is expressed as
$\sigma_{qed} \sim (\alpha^2/m_e^4)^2 \omega^6 \vartheta^4$~\cite{XsecQED}.
Therefore, we realize that the low frequency photons in qps is the best system
to probe such a low-mass field.

\subsection{Momentum uncertainty in the quasi-parallel system}\label{sec4}
\begin{figure}
\begin{center}
\includegraphics[width=0.7\linewidth]{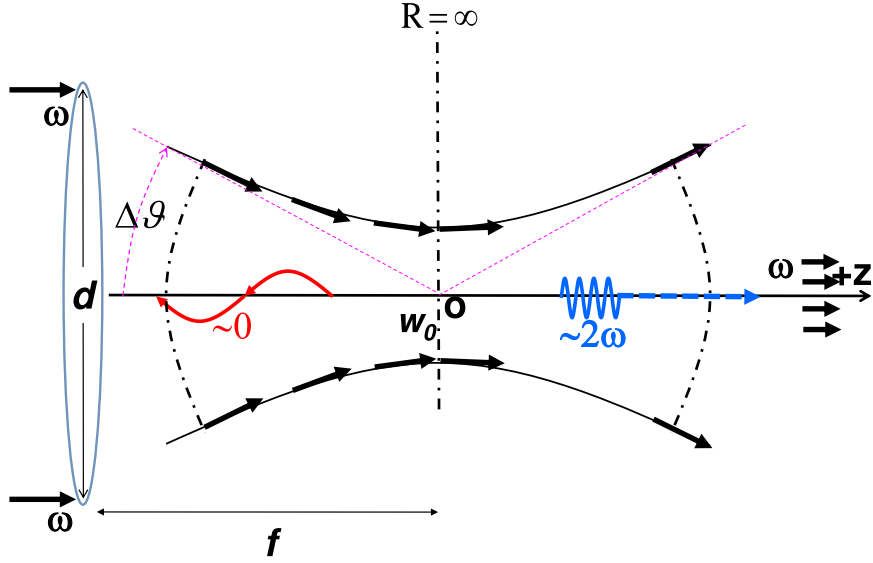}
\caption{
Frequency shift in the quasi parallel system by focusing a single
Gaussian laser beam defined with the incident frequency $\omega$,
the beam diameter $d$, the focal length $f$, and the beam waist $w_0$.
}
\label{Fig3}
\end{center}
\end{figure}

We note that it is not trivial to introduce two colliding photon beams
which satisfy the small incident angle based on the simple geometrical
optics due to the wavy nature of photons in the diffraction limit.
Below meV range we are naturally led to introduce a geometry by focusing
a single laser beam as illustrated in Fig.\ref{Fig3}.
In the diffraction limit there are uncertainties on the incident momentum 
due to the uncertainty principle. In other words there is uncertainty on the
incident angles between two photons among the single beam,
even though photons are in the degenerate state at the output of the laser
crystal. This should be contrasted to the case of a high energy collider where
the momentum spread of each colliding particle or the uncertainty based on
the de Broglie length is negligibly small compared to the magnitude of
relevant momentum exchanges of interest.
This different initial condition becomes critically important for 
the following discussion.

\subsection{How to overcome the extremely narrow resonance problem}\label{sec6}
The exact resonance condition is the requirement of $m = E_{cms}$ where
$m$ is the mass of exchanged field.
The square of the scattering amplitude $A$ can be expressed as
Breit-Wigner(BW) resonance function~\cite{BW}
\begin{equation}\label{eq_1}
|A|^2 = (4\pi)^2 \frac{W^2}{\chi^2(\vartheta) + W^2},
\end{equation}
where $\chi$ and the width $W$ are defined as
$\chi(\vartheta) \equiv \omega^2 - \omega^2_r(\vartheta)$ and
$W \equiv (\omega^2_r/16\pi)(g^2 m / M)^2$, respectively.
The energy $\omega_r$ satisfying the resonance condition can be
defined as $\omega_r \equiv m^2/(1-\cos 2\vartheta_r)$~\cite{DEapb}.
If we take $M$ as the Planckian mass, the width $a$ becomes extremely small.
This implies that the resonance width is too small to hit the peak position
of the resonance function. How can we overcome this problem?

We take a unique approach to this situation.
The Higgs hunting by high energy colliders is an example to hit the top
of the resonance.
In high energy colliders the spread of the beam energy
is much smaller than the width of the resonance function
to be probed. On the other hand, in our collisions at the diffraction limit
in qps, the resonance width is so tiny that it looks almost like
delta-function and the uncertainty on the $E_{cms}$ is much wider than 
the resonance width.
We may use the following feature of delta-function.
Although the width of the delta-function is
infinitesimal, as far as it is integrated over $\pm \infty$,
the value of integral becomes order of unity.
In the case of the BW function, even if
we integrate it over $\pm W$, the value of the integral is just a half of the
value integrated over $\pm \infty$. This implies that as long as
we capture the resonance peak within a finite range beyond $\pm W$,
the value of the integral becomes proportional to $a$ but not $W^2$.
In the diffraction limit the
incident angle of incoming photons is uncertain. Thus
we must use the averaged cross section by integrating the square of
the scattering amplitude over a possible range on $E_{cms}$ determined
by the uncertainty of the incident angle.

Let us reflect on this feature in the case of single beam focusing experiment.
We introduce the probability distribution function of the possible incident
angle between randomly selected photon pairs among the incident single laser
beam. We then define the range of integral on the incident angel as
$\Delta \vartheta \sim d/(2f)$ with the beam diameter $d$
and the focal length $f$.
If this range does not contain the resonance angle
$\vartheta_r \equiv m/2\omega$, that is, $\vartheta_r > \Delta \vartheta$,
the averaged squared amplitude
$|\overline{A}|^2$ becomes proportional to $a^2$, which indicates
the suppression by $M^{-4}$. On the other hand,
if $\vartheta_r < \Delta \vartheta$ is satisfied,
we obtain the proportionality to $a$, namely, a sensitivity enhancement
by $M^2$ compared to the case without resonance.
Thus various focusing parameters to adjust
$\Delta \vartheta$ by controlling the beam diameter and focal length
can introduce a sharp cutoff on the cross section
which eventually controls the sensitive mass range of this method
through the relation $m < 2\omega\Delta\vartheta$.

\subsection{Enhancement of interaction by the coherent states of high-intensity laser fields}
We now introduce luminosity in the laser collisions
in an analogy to the luminosity of the Fermionic particle colliders.
We define the time-integrated luminosity of the laser collision during
pulse duration time $\tau$ only in the vicinity of the focal point by
\beq
\int {\cal L} d\tau= \frac{{\cal I}}{\pi w^2_0},
\label{eqIntLumi}
\eeq
where $w_0$ is the beam-waist at the focal point. 
Here ${\cal I}=N^F(N^F-1)/2 \sim (N^F)^2$ 
is a dimensionless intensity corresponding to the combinatorics in
choosing pairs of incident particles out of the total number $N^F$
contained in a pulse, if the particles are Fermions with the 
distinguishable nature.  
For Bosonic beams, however, we face a difficulty to simply take 
combinatorics because a degenerate state of Bose particles such as 
coherent states are indistinguishable, hence uncountable. 
For this situation, we regard a laser pulse as the quantum coherent state 
which features a superposition of different photon numbers, characterized 
by an averaged photon number $N$ \cite{Glauber},
\beq
|N\dket \equiv \exp\left
(-N/2 \right) \sum_{n=0}^{\infty}
\frac{N^{n/2}}{\sqrt{n!}} |n>,
\label{eqCoherent0}
\eeq
where $|n>$ is the normalized state of $n$ photons
\beq
|n> =\frac{1}{\sqrt{n!}}\left( a^\dagger\right)^n |0>,
\label{eqCoherent1}
\eeq
with $a^\dagger$ and $a$  the creation and the annihilation
operators, respectively, of the photons assumed to share a single common
frequency and polarization. 
Although this is an approximation when 
applied to a pulse laser where multiple frequencies
must be included, it is a good starting point to investigate this 
simplified but well-defined approach.
The coherent state satisfies the normalization condition
\beq
\dbra N|N \dket=1.
\label{eqCoherent2}
\eeq
We also derive following properties of coherent states
$|N\dket$ and $\dbra N|$:
\beq
a |N\dket= \sqrt{N} |N\dket \mbox{ and }
\dbra N| a^{\dagger} = \sqrt{N} \dbra N|
\label{eqCoherent3}
\eeq
from the familiar relations
\beq
a^{\dagger}|n\rangle = \sqrt{n+1}|n+1\rangle \mbox{  and  }
a |n+1\rangle = \sqrt{n+1}|n+1\rangle.
\label{eqCoherent4}
\eeq
The property in Eq.(\ref{eqCoherent3}) gives the expectation value
of the annihilation and creation operators to coherent states
\beq
\dbra N|a|N \dket=\sqrt{N} \mbox{  and  }
\dbra N|a^{\dagger}|N \dket=\sqrt{N} \mbox.
\label{eqCoherent5}
\eeq

We now see how the coherent states are related with the Feynman amplitude
for the process $p_1 + p_2 \rightarrow p_3 + p_4$
by the exchange of a dark field with the mass $m$
as follows
\beq
 V_2\left( \left( p_1+p_2\right)^2+m \right)^{-1}V_1,
\label{6d_1}
\eeq
where $V_1$ and $V_2$ correspond to the matrix elements of the
interaction defined in Eq.(\ref{eqScalar}) or (\ref{eqPscalar})
at the first and the second vertices, respectively;
\beq
V_1 =gM^{-1}\bra 0|F_{\mu\nu}|1_{p_1} \ket \bra 0|F^{\mu\nu} |1_{p_2} \ket
\quad\mbox{and}\quad 
V_2 =gM^{-1}\bra 1_{p_3}|F_{\mu\nu}|0 \ket \bra 1_{p_4}|F^{\mu\nu}|0 \ket,
\label{6d_2}
\eeq
where $| 1_{p_i} \ket$ with $i=1, 2$ denotes one-photon states
specified with the individual momentum $p_i$.
To grasp the essence, by suppressing all the complications arising
from the momenta and polarization vectors, we abbreviate the factors
included in the initial vertex as follows
\beq
\bra 0|F_{\mu\nu}|1_{p_i} \ket \rightarrow \bra 0|a| 1_{p_i} \ket =1
\label{enhrev1_12},
\eeq
where the second of Eq.(\ref{eqCoherent4}) is applied to $n=0$.
We now consider the case where $p_i$ is annihilated within the coherent
sea with the same momentum and polarization state, and 
the proceeding interaction still occurs in the presence of the coherent 
field. We then extend the initial state from the one-photon state 
$|1_{p_i}\rangle$ to the coherent states $|N_{p_i}\dket$
\beq
\dbra N_{p_i}|a |N_{p_i}\dket = \sqrt{N_{p_i}},
\label{enhrev1_3}
\eeq
where the vacuum state $\langle0|$ is replaced by $\dbra N_{p_i}|$.
This is precisely the result of the first of Eq.(\ref{eqCoherent5}).  
Therefore, the presence of the coherent states enhances the matrix
element $V_1$  by $\sqrt{N_{p_1}}\sqrt{N_{p_2}}$.  
As for the second vertex $V_2$, the situation is different.
The outgoing photons with the momenta $p_3$ and $p_4$ are
quite different from the initial ones, because we require $p_3$ to be  
the observational signal with frequency up-shift. 
Thus, the final state photons are created spontaneously 
from the vacuum state $|0\rangle$ with no enhancement, 
because the sea of the coherent state specified with $p_3$ and $p_4$
does not exist. If $p_1 = p_2$ is realized by assuming that the incident 
two photons are annihilated in a single-frequency beam with the averaged 
number of photons $N$, the interaction rate is proportinal to 
the square of the Feynman amplitude. Therefore, we obtain
the enhancement factor $(\sqrt{N}\sqrt{N})^2$.
We associate the enhancement factor originating from the coherent state
with the dimensionless intensity in the time-integrated luminosity in
Eq.(\ref{eqIntLumi}). In this case, we see the correspondence 
${\cal I} \sim 1 \times (\sqrt{N}\sqrt{N})^2$ 
where $1$ explicitly expresses the combinatorics 
to take a virtual pair of laser pulses within a pulse.
We stress that this $N^2$ dependence for a laser pulse
in the coherent state accidentally coincides with the combinatorics 
factor $(N^F)^2$ for a pair of colliding Fermionic beam bunches.

Up to here, the coherent nature of the incident laser beam does not
improve the dimensionless intensity in the time-integrated luminosity
compared with that of the Fermionic particle collider. However,
if we could supply $|N_{p_4}\dket$ in advance of the interaction,
the situation drastically changes.
This additional laser beam, to be called an {\em inducing} beam, 
provides us with a sea of photons from which the photon $p_4$ is 
created in the induced manner,
so that $\bra p_4 |F^{\mu\nu} | 0\ket$ in the second of
Eq.(\ref{6d_2}) will be modified to
\beq
\bra 1_{p_4} |a^\dagger |0\ket = 1 \rightarrow 
\dbra N_{p_4}|a^\dagger |N_{p_4}\dket =\sqrt{N_{p_4}},
\label{6d_4}
\eeq
precisely the time-reversed process of the ones for the incident photons
with $p_1$ and $p_2$ in $V_1$. We note that photon $p_3$ remains to 
be created spontaneously from the vacuum because we must be able
to detect it as the signature of the interaction, 
while no attempt is made to observe the photon $p_4$, which will be embedded
quietly in $\dbra N_{p_4}|$.
We finally reach an important consequence; that is
the overall dimensionless intensity in the time-integrated luminosity is, 
therefore, expressed as
\beq\label{eqCubic}
{\cal I} = (\sqrt{N_{p_1}}\sqrt{N_{p_2}}\sqrt{N_{p_4}})^2 
= N_{p_1}N_{p_2}N_{p_3}.
\eeq
If $N_{p_i}$ is large enough, we have a huge dimensionless intensity compared 
with that of Fermionic particle colliders.

\subsection{Vacuum optical parametric effect (four-wave mixing)}
\begin{figure}
\begin{center}
\includegraphics[width=0.8\linewidth]{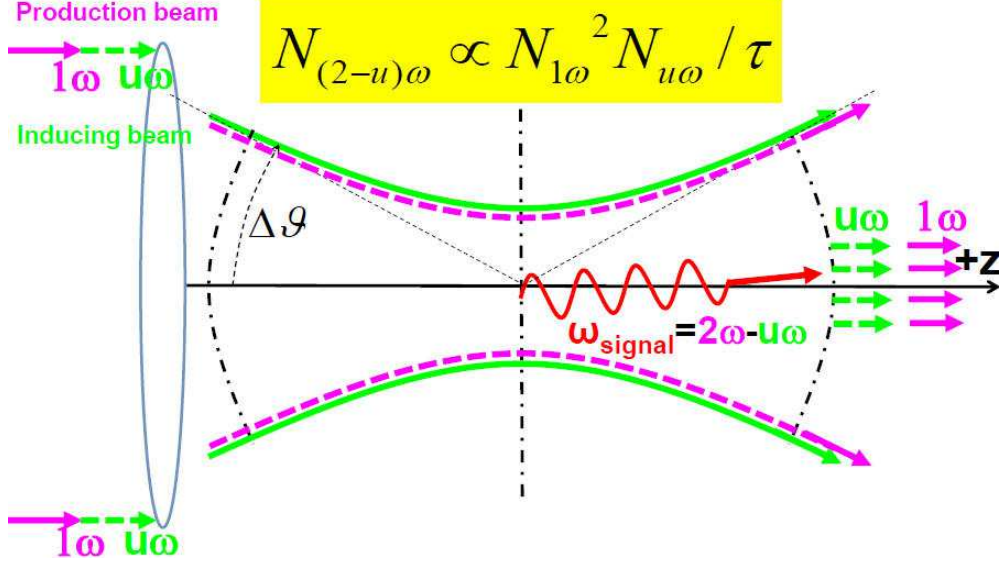}
\caption{
Degenerate four-wave mixing process in the vacuum.
}
\label{Fig4}
\end{center}
\end{figure}

In order to supply the inducing laser field, we propose to mix two laser 
fields with different frequencies $\omega$ and $u\omega$ where $u$ 
satisfies $0<u<1$ and $\omega$ is the fundamental laser frequency\cite{DEapb} 
as illustrated in Fig.\ref{Fig4}.  
The resonant interaction via the low-mass field exchange causes frequency 
shift by the process $\omega+\omega \rightarrow u\omega + (2-u)\omega$, 
where $u\omega$ acts as the inducing beam to make the low-mass field 
immediately decay and $(2-u)\omega$ is the signature that we measure as 
the indication of the photon-photon interaction. In quantum optics a 
similar process can occur which is known as the degenerate four-wave mixing 
(DFWM)\cite{Sylvie}. For example, 
$2\omega+2\omega \rightarrow 1\omega + 3\omega$ 
which corresponds to the case of $u=1/2$ with respect to the fundamental 
frequency $2\omega$. It is interesting to see the correspondence between 
the interaction in vacuum and that in nonlinear crystals despite of the 
difference in the fundamental processes. This process may correspond to the 
third susceptibility $\chi_3$ in Eq.(\ref{eq1}) from the observational point
of view. The coherent nature of laser fields enhances the interaction rate 
as we discussed above. In the case of DFWM, the interaction 
rate is expected to be proportional to the triple product of intensities 
of laser beams, because annihilation operator must be applied twice in 
the production vertex of low-mass fields and creation operator must be 
applied once in the decay vertex, if these three photons are annihilated 
or created under coherent states\cite{DEptp,Glauber}. We note that the 
inducing process never occurs in the vacuum, since only spontaneous decay 
is allowed in the vacuum state. The triple-product feature of DFWM can 
drastically increase the sensitivity to the weak coupling even beyond 
the gravitational coupling strength\cite{DEapb}, if the average number of 
photons per laser pulse is large enough, {\it e.g.}, if we could reach 
Avogadro's number corresponding to $\sim 200$ kJ. 

\subsection{High-Energy Colliders vs. High-Intensity Lasers}
Let us now compare the sensitivity of relevant event detection by the 
present boson-condensated beam collision with that by the conventional charged 
particle colliders. In most colliders, the number of charged particles per 
beam bunch is in the range of $10^{11}$ due to the space-charge limitation 
and other technical reasons. Therefore, the dimensionless intensity ${\cal I}$ 
in the numerator of the integrated luminosity over single beam bunch 
crossing time corresponds to $N^F_1N^F_2=10^{22}$, as we discussed
with Eq.(\ref{eqIntLumi}). When we collide the beam bunches at a repetition 
rate $\sim 1$GHz at most and taking data for three years, $10^8$ sec, 
the total dimensionless intensity is $10^{39}$ at most. 
On the other hand, in the proposed method, the corresponding 
Bosonic dimensionless intensity $\sim 20$ kJ per pulse focusing is 
$(10^{22})^3=10^{66}$. 
If we further assume a repetition rate 10kHz with lasers 
(the kind of technology to become available under the ICAN\cite{ICAN-HP} ) and 
a data taking period over, say, three years, the total dimensionless intensity
amounts to $10^{66} \times 10^4 \times 10^8 = 10^{78}$. This is greater 
than that of the conventional colliders by 39 orders of magnitude. 
This manifestly indicates the advantage of this method for the purpose of 
probing extremely small cross-sections and very weakly coupling interactions
when the necessary cms energy of the sought after fields is sufficiently low.

  We have seen the enhancement of the event detection by arranging the 
interaction of the coherent beams of laser fields in contrast to the 
incoherent collisions in the conventional charged particles beams. In this 
the coherent interaction is brought in via the coherence of laser fields.
Let us imagine that the fields we seek happen to exhibit Bosonic 
condensation (coherency) due to the nature of this field itself.  
If this is in fact the case, we might further experience collective 
phenomena via dynamics of unknown vacuum ingredients. For example, we know 
phonon resonance in normal matter where dynamics on the lattice itself 
causes the resonant phenomenon. If this kind of phenomenon could exist 
even in the vacuum, the proposed method is further distinct from high-energy 
colliders where the only point-like nature of vacuum can be investigated. 
The exploration of such possibility is embedded in our view as introduced 
in Eq.(\ref{eq0}). These considerations strongly motivate us to utilize the 
high-intensity lasers ultimately in the Zettawatt and Exawatt class for 
this suggestion\cite{HommaISMD}.

\section{Conclusion}
The advancement of high intensity ultrafast lasers as well as high photon
number (large energy) lasers greatly facilitates the enhanced reach of methods
of investigation of fundamental high energy physics. The current frontier of
electron acceleration by an ultrafast (fs) 1-10J capable of 
reaching energies of a GeV over a few cm is rapidly expanding.
We envision this frontier, in one rendition, expands toward 100 GeV to
TeV by adopting large energy lasers of 1-100 kJ
(a leap by some three orders of magnitude) to tune with the low-density
operating regime of the Laser Wakefield Acceleration over 10-100 m.
Although the science of laser acceleration is firmly grasped, there remain
a host of experimental techniques adjusted to the lower density plasma with
higher laser energy operation. Such an experiment amounts to 
a proof-of-principle demonstration of laser acceleration in the energy regime
of the Higgs boson mass. However, if we aspire to build a collider, we need
at least two more critical conditions, {\it i.e.} the repetition of such
accelerated lepton pulses to kHz and beyond 
(a leap by many orders of magnitude over what is available now) and the
efficiency of laser to tens of \% (another leap by a couple of orders of
magnitude). Fortunately, this formidable technological aspect is identified
(Leemans et al.\cite{86a}) under the collaborative work between the accelerator
community (represented by ICUIL, the International Committee for Ultahigh
Intensity Lasers), and more importantly is now given a possible route for
solution. The proposed technique of CAN (Coherent Amplification Network
\cite{Mourou2012}) adopts the coherent addition of highly efficient
fiber lasers. An international project has been launched (ICAN\cite{ICAN-HP}).
It is encouraging that over a relatively short period of time some important
building blocks toward this goals have been demonstrated, though details are
beyond the scope of this paper.

Although the aspiration of building a collider at the energy beyond the Higgs
mass is noble, we also suggest that there are more than one way to explore
fundamental physics with high energies\cite{PTP2011,Caldwell2012}. Some of
the physics issues we suggested do not require high luminosity (or does
call for only little luminosity), such as a test of relativity in very
high energies that may not be easy with the conventional accelerators but
may be reachable with laser plasma acceleration. This particular test may be
regarded as a part of examining the susceptibility of the vacuum as expanded 
by its nonlinearity and try to measure the leading term (linear term)
of Eq.(\ref{eq0}).

In addition to the high-energy research driven by laser acceleration, 
intense and large-photon beams of lasers can explore the property of 
the vacuum directly. The gateway to this task is to explore the QED vacuum
nonlinearity in its susceptibility induced by intense laser field.
This intense pump field nonlinearly induces polarization of the vacuum, 
which may be probed by the probe pulse of laser. We have suggested a 
particular version of phase contrast imaging of the vacuum nonlinearity 
based on QED. There is large and growing literature on this and related 
subject in recent years, spurred by the prospect of near-term possibility 
to be able to detect such QED nonlinearities.

On one hand it is exciting to measure and detect nonlinear phenomena of
the vacuum induced and controlled by intense laser field. 
This is a bit similar to the situation that was created right after the 
invention of laser in 1960 to spawn out nonlinear optics in 1961 
(2nd harmonic generation in quartz by P. Franken\cite{Franken}), when
atomic physics emerged as a far more refined discipline with far accurate
controllability and broader applicability. Because Heisenberg-Euler's
Lagrangian has the precise ratio 4:7 on scalar vs. pseudoscalar contributions,
if and when the measured branching ratio deviates from this, it implies
that some new unknown fields are lurking behind QED fields.
In order to enhance the detectability of very weakly coupling fields
with photons and thus so far undiscovered field (thus called 'dark'),
we have introduced the four-wave mixing method of vacuum search.
This again has a parallel to the nonlinear optics in media. Some of the
theoretical candidates for Dark Matter and Dark Energy are extremely weak
coupling fields with very low-mass for the mediating field particles
(such as meV-$\mu$eV for axion-like particles and neV for a Dark Energy 
candidate). These particle-fields are so light in contrast to the present
day mass of particles at the frontier of HEP. It is thus far better match
for mass-less photons as the search engine rather than extreme high
energy particle beams of a collider. Our proposed approach of using near
co-parallel propagation of laser pulses as well as our resonant interaction
nature of four-wave mixing allow us to hugely enhance the sensitivity.
As compared with the colllider of particles where the luminosity per shot 
is proportional to $N^F_1N^F_2$ in Eq.(\ref{eq1}), the leaser four-wave
mixing luminosity per shot is proportional to $N_{p_1}N_{p_2}N_{p_3}$ 
in Eq.(\ref{eqCubic}).
We have seen a potential enhancement of sensitivity by a huge factor of
$10^{30-40}$. In order to make such a method work, we need to minimize the
noise due to the material responses, instrumental errors etc., while maximize 
the sought after signals.

\begin{figure}
\begin{center}
\includegraphics[width=1.0\linewidth]{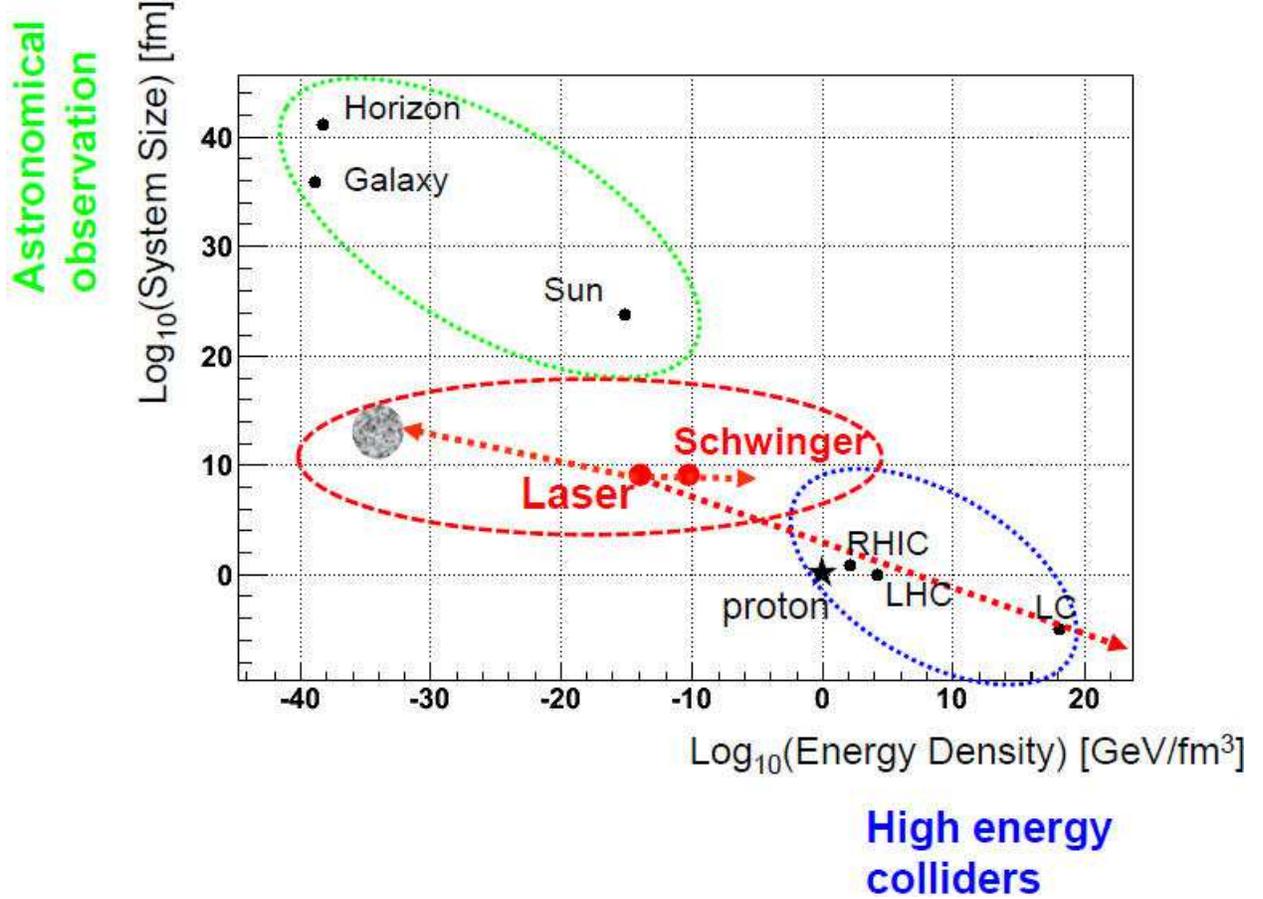}
\caption{
Three directions of research extended by high-intensity and large-energy
laser fields in particle physics and cosmology in the future.
Domain sizes where physical laws appear or materialize are plotted as a 
function of the energy densities relevant to the laws. The proton is taken 
to be the origin of the plot. Astronomical objects and collision energies 
per particle volumes at high energy particles colliders,
AA(RHIC), pp(LHC) and $e^+e^-$(LC), are mapped on the figure~\cite{HommaISMD}.
}
\label{Fig5}
\end{center}
\end{figure}

As a concluding remark, we summarize the proposed approaches 
by high-intensity as well as large-energy laser fields.
The directions discussed by this paper are mapped as the three arrows
on Fig.\ref{Fig5}, which illustrate the domains specified with the sizes and
energy densities being the proton as the origin of the plot~\cite{HommaISMD}. 
Various kinds of physical laws appear or materialize depending on the domain. 
The experimental approaches to probe relevant phenomena to the physical 
laws are also indicated (see the caption of the figure in detail).
LWFA combined with ICAN technology extends the arrow to the microscopic
high-energy density domains or digging the vacuum by point-like probes.
The new field search of the ether-like view point enriches the insight into the nature of the vacuum
via the order parameter such as susceptibility of the vacuum polarization
with respect to external intense fields, as defined in Eq.(\ref{eq0}). 
This view point naturally leads us to the remaining two directions.
The domain pointed by the arrow right to the horizontal direction would be explorable by the
well-controlled high-intensity lasers combined with energetic particles,
in which we can probe non-perturbative aspects of the vacuum structure
via the real and imaginary parts of the vacuum polarizations relevant to
$\chi_2$. This is rather difficult to be tested by conventional high-energy 
colliders in a controlled matter because of the extremely short interaction 
time scale as seen in the early stage of the quark-gluon plasma formation
in high-energy heavy-ion collisions. In addition, given high-energy photons
by the direction of the first arrow, we may explore whether $\chi_1$,
relevant to the velocity of light, is really constant or not over a wide 
range of frequencies of probe photons.
The remaining domain pointed by the arrow left to low energy density and 
semi-macroscopic distance may explore the light dark components of the 
universe as a collection 
of the local vacuum structure with feeble couplings to ordinary 
matter. Large-energy laser fields properly arranged for the four-wave mixing 
configuration relevant to accentuate $\chi_3$ may provide a new approach 
to probe the long-range structure of the vacuum.
Since the sizes of the accessible domains are very different from what are
explored by conventional methods in particle physics and cosmology, 
the proposed approaches potentially provide new bridges 
between particle physics and cosmology in many aspects.

\section*{Acknowledgments}
We thank
G. Mourou, S. Bulanov, T. Esirkepov, Y. Fujii, M. Kando, P. Chen, F. Krausz, 
E. Goulielmakis, N. Narozhny, A. Chao, B. LeGarrec, D. Jaroszynski, 
M. Spiro, A. Suzuki, D. Habs, W. Leemans, K. Nakajima, D. Payne, 
X. Yan, A. Caldwell, R. Heuer, P. Bolton, A. Ipp, H. Ruhl, C. Klier, 
J. Nilsson, C. Cohen-Tannoudji, and E. Esarey for their discussions,
contributions, and encouragements.
This work was supported by the Grant-in-Aid for Scientific
Research no.24654069 from MEXT of Japan in part.

\newpage


\begin{thebibliography}{99}
\bibitem{MTigner}
M. Tigner, Nuovo Cimento 37, 1228 (1965).
\bibitem{Keldysh}
L. Keldysh et al. Sov. Phys. JETP 20, 1307 (1965).
\bibitem{TT-Dawson}
T. Tajima and J. Dawson, Phy. Rev. Lett. 43, 267 (1979).
\bibitem{Chen}
P. Chen et al., Phys. Rev. Lett. 56, 1252 (1986).
\bibitem{Caldwell}
A. Caldwell et al., Nature Phys. 5, 363 (2009).
\bibitem{Zheng2012}
F. L. Zheng et al., Phys. Plasma 19, 023111 (2012).
\bibitem{Chao-Tigner}
A. Chao and M. Tigner, "Handbook of Accelerator Physics and Engineering",
(World Scientific, Singapore, 1999).
\bibitem{SLAC-lec.-2008}
A. Suzuki, SLAC ICFA lecture (SLAC, Stanford, 2008).
\bibitem{Veksler}
V. Veksler, Proc. CERN Symp. 1, 68 (1956).
\bibitem{Rostoker-Reiser}
N. Rostoker and M. Reiser, "Collective Methods for Acceleration",
(Harwood, Amsterdam, 1979).
\bibitem{Esirkepov2004}
T. Esirkepov, M. Borghesi, S.V. Bulanov, G. Mourou, and T. Tajima,
Phys. Rev. Lett. 92, 175003 (2004).
\bibitem{Snavely2000}
R. Snavely et al. Phys. Rev. Lett. 85, 2945 (2000).
\bibitem{Mourou2012}
Mourou, G., Fisch, N., Malkin, V.M. Toroker, Z., Khazanov, E. A., Sergeev, 
A. M., Tajima, T., and Le Garrec, B., Opt. Comm. 285, 720 (2012).
\bibitem{Narozhny2004}
N. Narozhny et al., Phys. Lett. A 330, 1 (2004).
\bibitem{Tajima2010}
T. Tajima, Laser  Part. Beams 3, 351 (1985).
\bibitem{Cohen-Tannoudji}
C. Cohen-Tannoudji and D. Guery-Odelin, 
"Advances In Atomic Physics: An Overview", (World Scientific, Singapore, 1999).
\bibitem{Tajima2011}
T. Tajima, PASJ B 86, 147 (2011).
\bibitem{Pirozhkov2012}
A. Pirozhkov et al., Phys. Rev. Lett. 108, 135004 (1985).
\bibitem{Esarey}
E. Esarey et al. IEEE Trans. Plas. Sci. 24, 252 (1996).
\bibitem{Schroeder}
C. Schroeder et al., Phys. Rev. STAB 13, 101301 (2010).
\bibitem{Nakajima2011}
K. Nakajima, A. Deng, X. M. Zhang, B.F. Shen, J. S. Liu, R. X. Li, Z. Z. Xu, T. Ostermayr, S. Petrovics, C. Klier, K. Iqbal, H. Ruhl, and T. Tajima, Phys. Rev. STAB 14, 09130 (2011).
\bibitem{Khalid2012}
K. Nakajima et al., Phys. Rev. STAB 15, 081303 (2012);
I. Kostyukov et al., submitted to Phys. Rev. STAB(2012).
\bibitem{PTP2011}
T. Tajima, M. Kando, and M. Teshima, Prog. Theor. Phys. 125, 617 (2011).
\bibitem{Nakajima95}
Nakajima, K., Fisher, D., Kawakubo, T., Nakanishi, H., Ogata, A., Kato, Y., Kitagawa, Y., Kodama, R., Mima, K., Shiraga, H., Suzuki, K., Yamakawa, K., Zhang, T., Sakawa, Y., Shoji, T., Nishida, Y., Yugami, N., Downer, M. and Tajima, T., Phys. Rev. Lett. 74, 4428 (1995).
\bibitem{Madena95}
A. Madena et al.,  Nature, 377, 606 (1995).
\bibitem{Faure}
J. Faure , Y. Glinec , A. Pukhov , S. Kiselev , S. Gordienko , E. Lefebvre , J.-P. Rousseau , F. Burgy and V. Malka, Nature 431, 541 (2004).
\bibitem{Mangle}
S. Mangle et al., Nature 431, 535 (2004);
C. Geddes, et al. Nature 431, 538 (2004).
\bibitem{Leemans2006}
W. Leemans et al., Nature Phys. 2 696 (2006).
\bibitem{Xie1997}
Xie, M., Tajima, T., Yokoya, K. and Chattopadyay, S., (AIP Conference Proceedings, New York, 1997), 398, p. 233-242.
\bibitem{Esarey2009}
E. Esarey et al., Rev. Mod. Phys. 81, 1229 (2009).
\bibitem{IZET-HP}
www.izest.polytechnique.edu
\bibitem{ICFA-Newslett-2011}
W. Leemans, M. Uesaka, W. Chou, eds, ICFA Newsletter 50, 10 (2011);
G. Mourou, D. Hulin, and A. Galvanauskas, AIP Conf. Proc. 827, 152 (2006).
\bibitem{ICAN-HP}
https://www.izest.polytechnique.edu/izest-home/ican/ican-94447.kjsp?RF=1332339530225
\bibitem{PRL19}
Z. Huang, P. Chen, and R. Ruth, Phys. Rev. Lett. 74, 1759 (1995).
\bibitem{Ashour-Abdalla1981}
M. Ashour-Abdalla, M., Leboeuf, J.N., Tajima, T., Dawson, J.M. and Kennel, C.F., Phys. Rev. A, 23, 1906 (1981).
\bibitem{Esirkepov2002}
Esirkepov, T.Zh., Bulanov, S.V., Nishihara, K., Tajima, T., Pegoraro, F., Khoroshkov, V.S., Mima, K., Daido, H., Kato, Y., Kitagawa, Y., Nagai, K., and Sakabe, S., Phys. Rev. Lett.89,175003(2002)
\bibitem{Bulanov2003}
S. Bulanov et al. Phys. Rev. Lett 91, 085001 (2003).
\bibitem{Kando}
Kando, M., Fukuda, Y., Pirozhkov, A.S., Ma, J., Daito, I., Chen, L.M., Esirkepov, T.Z., Ogura, K., Homma, T., Hayashi, Y., Kotaki, H., Sagisaka, A., Mori, M., Koga, J.K., Daido, H., Bulanov, S.V., Kimura, T., Kato, Y., Tajima, T., Phys.Rev.Lett.99, 135001 (2007).
\bibitem{Pirozhkov}
Pirozhkov, A.S., Ma, J., Kando, M., Esirkepov, T., Fukuda, Y., Chen, L.M., Daito, I., Ogura, K., Homma, T., Hayashi, Y., Kotaki, H., Sagisaka, A., Mori, M., Koga, J.K., Kawachi, T., Daido, H., Bulanov, S.V., Kimura, T., Kato, Y., Tajima, T., Phys. Plasmas 14, 123106 (2007).
\bibitem{Ipp2011}
A. Ipp et al., Phys. Lett. B 702, 383 (2011).
\bibitem{Tajima2012}
in preparation
\bibitem{Nikishov-Ritsu}
A. Nikishov and V. Ritus, Sov. Phys. JETP 19, 529 (1964).
\bibitem{Goulielmakis2008}
E. Goulielmakis et al., Science 317, 769 (2008).
\bibitem{Sato}
T. Sato et al., Prog. Theor. Phys. 47, 1788 (1972).
\bibitem{Sidharth}
B. Sidharth, "Discrete Space-Time and Lorentz Symmetry", 
(Springer, Berlin, 2004).
\bibitem{Ellis}
G. Amelio-Camelia, "Anything beyond Special Relativity",
(Springer, Berlin, 2006).
\bibitem{Abdo}
A. Abdo et al., Nature 462, 33 (2008);
\bibitem{LeGarrec}
J. Di-Nicolis et al., J. Phys. IV France, 113, 595 (2006).
\bibitem{Ostermayr}
T. Ostermayr, S. Petrovics, K. Iqbal, C. Klier, T. Tajima, H.  Ruhl, K. Nakajima, A. Deng, X. M.  Zhang, B. F. Shen, J. S. Liu, R. X. Li, and Z. Z. Xu, accepted by J. Plasma Phys. (2012). 
\bibitem{Altschul}
N. Altschul, Phys. Rev. D80, 091901 (2009).
\bibitem{Dino}
S. Cipiccia, et al., Nature Phys. 7, 867 (2011).
\bibitem{Schwinger}
J. Schwinger, Phys. Rev. 82, 664 (1951).
\bibitem{Narozhny}
N. Narozhny, JTEP 27, 360 (1968).
\bibitem{Ruhl}
N.V. Elkina, A.M. Fedotov, I.Yu. Kostyukov, M.V. Legkov, N.B. Narozhny, E.N. Nerush, H. Ruhl, Phys.Rev.ST Accel.Beams 14: 054401,(2011).
\bibitem{53a}
G. Mourou, T. Tajima, and S. Bulanov, Rev. Mod. Phys. 78, 309 (2006).
\bibitem{54a}
G. Mourou and T. Tajima, Science 331, 41 (2011).
\bibitem{Dirac}
P. A. M. Dirac, Proc. R. Soc. Lond. A 167, 148 (1938).
%
%
\bibitem{Shore}
G. M.~Shore,
Nucl.\ Phys.\  B {\bf 778}, 219 (2007).
\bibitem{KN}
R. Karplus and M. Neuman, Phys. Rev. {\bf 83} 776-784 (1950).
\bibitem{DT}
B. De Tollis, Nuovo Cimento {\bf 32} 757 (1964);
B. De Tollis, Nouvo Cimento {\bf 35} 1182 (1965).
\bibitem{APB-QED} 
K.Homma, D. Habs, and T. Tajima,
Applied Physics B Lasers and Optics (2011)104:769782 
(DOI:10.1007/s00340-011-4568-2), arXiv:1104.0994[hep-ph].
\bibitem{EH}
W.~Heisenberg and H.~Euler, Z.\ Phys.\  {\bf 98}, 714 (1936).
\bibitem{Weiscop}
V. Weisskopf, Kong. Dans. Vid. Selsk. Math-fys. Medd. {\bf XIV}, 166 (1936).
\bibitem{AxionReview}
See, for instance, the reviewing chapter of
J. Beringer et al. (Particle Data Group), Phys. Rev. D86, 010001 (2012).
\bibitem{ABJ}
S. Adler, Phys. Rev. 177, 2426 (1969);
J. S. Bell and R. Jackiew, Nuovo Cimento 60A, 47 (1967).
\bibitem{StrongCP}
R. D. Peccei, {\it The Strong CP Problem in "CP Violation"},
Advanced Series on Derections in High Energy Physics, Vol.3, ed.
C. Jarlskog, World Scientifc, Singapore (1989).
\bibitem{tHooft}
G. 't Hooft. Phys. Rev. Lett. 37, 8 (1976).
\bibitem{Higgs}
P. W. Higgs, Phys. Rev. Lett. 13, 508509 (1964).
\bibitem{NeutronDipole}
C. A. Baker et al., Phys. Rev. Lett. 97, 131801 (2006).
\bibitem{PQ}
R. D. Peccei and H. R. Quinn,
Phys. Rev. Lett. {\bf 38}, 1440 (1977).
\bibitem{Weinberg}
S. Weinberg, Phys. Rev. Lett. 40, 223 (1978).
\bibitem{Wilczek} 
F. Wilczek, Phys. Rev. Lett. 40, 271 (1978).
\bibitem{AxionCosmology}
E. W. Kolb and M. S. Turner, {\it The Early Universe}, 
Westview Press, Boulder, Colorado (1990).
\bibitem{isocurvature}
Mark P. Hertzberg, Max Tegmark, and Frank Wilczek,
Phys. Rev. D {\bf 78}, 083507 (2008).
\bibitem{DiracLargeNumber}
P. A. M. Dirac, Proc. R. Soc. Lond. A 165, 199-208 (1938).
\bibitem{Jordan}
P. Jordan, {\it Schwerkraft und Weltall} (Friedrich Vieweg und Sohn, Brunschweig, 1955).
\bibitem{STTL}
Y. Fujii and K. Maeda, {\it The Scalar-Tensor Theory of Gravitation}
(Cambridge Univ. Press, 2003).
\bibitem{STRING}
Eq.(3.4.58) in 
M. B. Green, J. H. Schwarz and E. Witten, {\it Superstring Theory},
Cambridge University Pres (1985).
\bibitem{BD}
C. Brans and R. H. Dicke, Phys. Rev. {\bf 124} (1961), 925.
\bibitem{nat}
Y. Fujii, Nature Phys. Sci. {\bf 234} (1971), 5; Ann. of Phys. {\bf 69} (1972), 494.
\bibitem{AxionCDM}
Mark P. Hertzberg, Max Tegmark, and Frank Wilczek,
Phys. Rev. D {\bf 78}, 083507 (2008);
O. Wantz and E. P. S. Shellard, Phys. Rev. D 82, 123508 (2010).
\bibitem{DEptp}
Y. Fujii and K. Homma,
Prog. Theor. Phys. 126: 531-553 (2011), arXiv:1006.1762 [gr-qc].
\bibitem{WEPV}
See, for example, Figures; 2.13, 4.16-17 in E. Fischbach and
C. Talmadge, {\it The Search for Non-Newtonian Gravity} (AIP Press,
Springer-Verlag, New York, 1998).  \\
See also S. Schlamminger, K.-Y. Choi, T. A. Wagner, J. H. Gundlach and E. G. Adelberger, Phys. Rev. Lett. {\bf 100} (2008), 041101, and papers cited therein.
\bibitem{DEapb}
K. Homma, D. Habs, T. Tajima,
Appl. Phys. B 106:229-240 (2012),
(DOI: 10.1007/s00340-011-4567-3),arXiv:1103.1748 [hep-ph].
\bibitem{Franken}
P. Franken, A. E. Hill, C. W. Peters, and G. Weinreich,
Phys. Rev. Lett. {\bf 7}, 118 (1961).
\bibitem{XsecQED}
See p.183 in W.~Dittrich and H.~Gies,
{\it Probing the Quantum Vacuum} (Springer, Berlin, 2007).
\bibitem{BW}
For example, see section for 
{\it Cross-section formulae for specific processes}
in C. Amsler et al. (Particle Data Group),
Phy. Lett. B{\bf 667}, 1 (2008) and 2009 partial update for the 2010 edition.
\bibitem{Glauber}
R. J. Glauber, Phys. Rev. {\bf 131} (1963), 2766.
\bibitem{Sylvie}
Sylvie A. J. Druet and Jean-Pierre E. Taran,
Prog. Quant. Electr. Vol.7, pp. 1-72 (1981).
\bibitem{HommaISMD}
K. Homma, D. Habs, G. Mourou, H. Ruhl, and T. Tajima,
Prog. Theor. Phys. Suppl. No. 193, (2012);
http://www.extreme-light-infrastructure.eu/;
http://www.int-zest.com/index.html.
\bibitem{86a}
W. Leemans et al., in Beam Dyanamics Newsletter 56, 10  (2011), eds. W. Chou et al.
\bibitem{Caldwell2012}
G. Dvali, G. F. Giudice, C. Gomez and A. Kehagias, JHEP 1108, 108 (2011).
\end{thebibliography}
\end{document}